\documentclass[a4paper,11pt]{article}
\pdfoutput=1
\usepackage{jheppub}
\usepackage{graphicx,slashed,hyperref,color,multirow}
\usepackage{amssymb}
\usepackage[normalem]{ulem}
\usepackage[utf8]{inputenc}
\usepackage{braket}
\usepackage{pifont}
\hypersetup{
   bookmarks=true,         
   unicode=true,          
   pdftoolbar=true,        
   pdfmenubar=true,        
   pdffitwindow=false,     
   pdfstartview={FitH},    
   pdftitle={My title},    
   pdfauthor={Author},     
   pdfsubject={Subject},   
   pdfcreator={Creator},   
   pdfproducer={Producer}, 
   pdfkeywords={keyword1} {key2} {key3}, 
   pdfnewwindow=true,      
   colorlinks=true,       
   linkcolor=blue,          
   citecolor=magenta,        
   filecolor=magenta,      
   urlcolor=cyan           
}

\newcommand{\cmark}{{\color{green} \ding{51}}}%
\newcommand{\xmark}{{\color{red} \ding{55}}}%

\def\lsim{\mathrel{\rlap{\lower4pt\hbox{\hskip1pt$\sim$}}
    \raise1pt\hbox{$<$}}}         
\def\gsim{\mathrel{\rlap{\lower4pt\hbox{\hskip1pt$\sim$}}
    \raise1pt\hbox{$>$}}}         

\makeatletter

\newcommand{\fmslash}[2][0mu]{%
  \mathchoice
    {\fmsl@sh\displaystyle{#1}{#2}}%
    {\fmsl@sh\textstyle{#1}{#2}}%
    {\fmsl@sh\scriptstyle{#1}{#2}}%
    {\fmsl@sh\scriptscriptstyle{#1}{#2}}}
\newcommand{\fmsl@sh}[3]{%
  \m@th\ooalign{$\hfil#1\mkern#2/\hfil$\crcr$#1#3$}}
\makeatother

\newcommand{\beq}{\begin{equation}}
\newcommand{\eeq}{\end{equation}}
\newcommand{\bea}{\begin{eqnarray}}
\newcommand{\eea}{\end{eqnarray}}
\mathchardef\minus="002D

\usepackage{color}

\def\beq{\begin{equation}}
\def\eeq{\end{equation}}
\def\bea{\begin{eqnarray}}
\def\eea{\end{eqnarray}}

\title{Searching for Dark Matter Signals in \\
Timing Spectra at Neutrino Experiments}

\author[a]{Bhaskar Dutta,}
\author[a]{Doojin Kim,}
\author[a]{Shu Liao,}
\author[b]{Jong-Chul Park,}
\author[c]{Seodong Shin,}
\author[a]{Louis E. Strigari,}
\author[a]{and Adrian Thompson}

\affiliation[a]{Mitchell Institute for Fundamental Physics and Astronomy, Department of Physics and Astronomy, Texas A\&M University, College Station, TX 77843, USA}
\affiliation[b]{Department of Physics and Institute of Quantum Systems (IQS), Chungnam National University, Daejeon 34134, Republic of Korea}
\affiliation[c]{Department of Physics, Jeonbuk National University, Jeonju, Jeonbuk 54896, Republic of Korea}

\emailAdd{dutta@physics.tamu.edu}
\emailAdd{doojin.kim@tamu.edu}
\emailAdd{ikaros@physics.tamu.edu}
\emailAdd{jcpark@cnu.ac.kr}
\emailAdd{sshin@jbnu.ac.kr}
\emailAdd{strigari@tamu.edu}
\emailAdd{thompson@physics.tamu.edu}

\preprint{
\begin{minipage}{5cm}
\begin{flushright}
MI-TH-2014
 \end{flushright}
\end{minipage}
}

\abstract{
The sensitivity to dark matter signals at neutrino experiments is fundamentally challenged by the neutrino rates, as they leave similar signatures in their detectors.
As a way to improve the signal sensitivity, we investigate a dark matter search strategy which utilizes the timing and energy spectra to discriminate dark matter from neutrino signals at low-energy, pulsed-beam neutrino experiments.
This strategy was proposed in our companion paper {\it PRL {\bf 124} (2020) 121802}~\cite{Dutta:2019nbn}, which we apply to potential searches at COHERENT, JSNS$^2$, and CCM.
These experiments are not only sources of neutrinos but also high intensity sources of photons.
The dark matter candidate of interest comes from the relatively prompt decay of a dark sector gauge boson which may replace a Standard-Model photon, so the delayed neutrino events can be suppressed by keeping prompt events only.
Furthermore, prompt neutrino events can be rejected by a cut in recoil energy spectra, as their incoming energy is relatively small and bounded from above while dark matter may deposit a sizable energy beyond it.
We apply the search strategy of imposing a combination of energy and timing cuts to the existing CsI and LAr data of the COHERENT experiment as concrete examples, and report a mild excess beyond known backgrounds.
We then investigate the expected sensitivity reaches to dark matter signals in our benchmark experiments. 
\newpage
}

\begin{document}

\maketitle

\section{Introduction \label{sec:introduction}}

While there exists a tremendous amount of astrophysical and cosmological evidence for the existence of dark matter in the universe, none of the experimental efforts including dark matter direct/indirect detection experiments and accelerator-based experiments have made conclusive observations via its hypothetical non-gravitational interactions with Standard Model (SM) particles. 
In particular, null results in the search for Weakly Interacting Massive Particles~\cite{Aprile:2018dbl}, which have been considered as a well-motivated dark matter candidate, have stimulated searches for dark matter candidates dwelling in other mass regions~\cite{Battaglieri:2017aum}.
Among the possible candidates, MeV to sub-GeV-range dark matter in connection with portal scenarios have received particular attention, as it is (\textit{i}) a kind of thermal dark matter, (\textit{ii}) relatively less constrained by existing bounds, and (\textit{iii}) readily explorable in many ongoing and projected high-intensity beam-based experiments. 
There are already many interesting physics models addressing this class of light dark matter, especially in scenarios with light mediators or portal particles (e.g., dark photons) interacting with the dark matter (see, for example, Refs.~\cite{Huh:2007zw, Pospelov:2007mp, Hooper:2008im, Cheung:2009qd, Essig:2010ye, Essig:2013lka,Batell:2014yra, Dutta:2019fxn}). 

Searches for dark matter candidates at neutrino experiments using high intensity, $\mathcal{O}(1)$ GeV particle beams are particularly intriguing\cite{deNiverville:2011it,deNiverville:2015mwa,Ge:2017mcq,Dutta:2019nbn,Akimov:2019xdj,Berlin:2020uwy}.
The weak couplings between the SM and dark sectors through portal particles can be compensated by high statistics.
Laboratory-based searches can also benefit from better handles to control potential backgrounds and environmental parameters, in contrast with the astrophysical searches.
Despite these advantages, however, neutrinos can mimic dark matter signals by having similar experimental signatures. 
As a consequence, maximizing the dark matter signal sensitivity is hampered by irreducible neutrino-induced contamination to the dark matter signal, so it is highly desired to devise an intelligent search strategy to get around this issue. 

An earlier effort was made to utilize the energy cut to reduce certain neutrino backgrounds whose energy deposit at the detector is kinematically bounded from above. 
Example studies include the dark matter searches at the LSND and MiniBooNE detectors~\cite{deNiverville:2011it}, at the COHERENT and CENNS -- which is now part of the COHERENT experiment -- detectors~\cite{deNiverville:2015mwa}, and at the COHERENT and CCM detectors~\cite{Berlin:2020uwy}. 
As we will discuss shortly, however, this does not suffice to suppress other neutrino backgrounds while retaining reasonable signal statistics. 
By contrast, it was realized that adoption of a timing cut is useful to reject the neutrino events lying in delayed timing bins, e.g., COHERENT~\cite{Dutta:2019nbn,Akimov:2019xdj}.
In particular, Ref.~\cite{Akimov:2019xdj} makes use of the delayed timing bins to determine the spectral behavior of neutrino events in the prompt timing bins up to statistical and systematic errors through a side-band analysis. 
However, this approach may be fully valid in the absence of non-standard interactions (NSIs) of neutrinos, potentially introducing some model dependencies, and the associated signal region may suffer from a sizable number of neutrino background events, essentially limiting the signal sensitivity.  

In light of this situation, we have pointed out that timing along with energy selection cuts can enhance the signal sensitivity in new physics searches by vetoing neutrino backgrounds efficiently and model-independently, in our companion paper~\cite{Dutta:2019nbn}.
We remark that many of the intensity-frontier neutrino experiments come with an apparatus and/or an algorithm to record the timing information of an event occurrence at a detector. 
In a low-energy neutrino experiment, a particle beam impinges on a target, producing neutrinos through the decays of mesons (mostly charged pions) and muons at rest. 
There are two classes of neutrinos, ``prompt'' neutrinos from the decay of (stopped) $\pi^\pm$ and ``delayed'' neutrinos from the decay of relatively longer-lived (stopped) $\mu^\pm$. 
The dark matter candidate of our interest is created by a rather prompt decay of a dark sector gauge boson, and thus signal events preferentially populate in prompt timing bins. 
Therefore, it is possible to suppress the delayed neutrino events significantly by selecting the events populating the prompt timing bins. 
By contrast, the energy of the prompt neutrinos is single-valued as they come from the two-body decay of stopped $\pi^\pm$ in association with a massive muon.
Since most of the $\pi^\pm$ (mass) energy is carried away by the muon, the energy deposit allowed for prompt neutrinos is not large.
On the other hand, typical dark matter events can deposit larger energies.
So, an energy cut can allow to veto prompt neutrino events very efficiently while keeping a large fraction of the dark matter events. 

While many of the low-energy, high-intensity frontier experiments possibly would enjoy the advantage of a combination of energy and timing cuts in terms of dark matter searches, we particularly focus on the COHERENT~\cite{Akimov:2017ade, Akimov:2018ghi, Akimov:2019xdj}, JSNS$^2$~\cite{Ajimura:2015yux, Ajimura:2017fld, privateJSNS2}, and CCM~\cite{CCM, dunton_2019, privateCCM} experiments to apply our search techniques for. 
They are equipped with an $\mathcal{O}(1~{\rm GeV})$ beam delivering $\sim 10^{22}-10^{23}$ protons-on-target (POT) per year, hence producing dark matter particles copiously.
Furthermore, most of their projected detectors are featured by a tonne scale in mass, so a large amount of exposure to the dark matter flux is expected even with a small amount of duty time. 

As mentioned earlier, dark matter production is initiated by production of a portal particle which mediates the interactions between SM particles and dark matter. 
In this study, we consider vector portal-type scenarios whose mediators are henceforth collectively called dark(-sector) gauge bosons, while the generic strategy is readily applicable to other portal scenarios (e.g., scalar portal).
The protons struck on a target often induce creation of light mesons (mostly $\pi^0$, $\pi^\pm$) which subsequently involve photon production: for example, $\pi^0$ decay ($\pi^0\to 2\gamma$) and $\pi^-$ absorption ($\pi^- +p\to n+\gamma$). 
A photon in these processes may be replaced by a dark gauge boson whose coupling to the SM particles is sufficiently small to satisfy existing limits. 
The dark gauge boson then disintegrates to dark matter particles some of which may freely stream toward a detector and leave a signature via nucleus or electron recoil in the detector material. 
It is crucial to estimate the differential photon flux and pion flux (for the $\pi^-$ absorption case) in the target as accurately as possible. 
To this end, we employ the \texttt{GEANT}4 10.5 (FTFP\_BERT) code package~\cite{Agostinelli:2002hh} to take into account detailed nuclear effects. By doing so, additional photon sources such as $e^\pm$-induced cascade photons are included, and we find that their contribution can be non-negligible depending on the mass of the dark gauge boson. 

To illustrate how the proposed search strategy is realized in the actual data analyses, we explicate the measurement data collected at the CsI and liquid argon (LAr) detectors of the COHERENT experiment~\cite{Akimov:2018vzs,Akimov:2020pdx}.
Our analysis scheme with an appropriately chosen set of energy and timing cuts allow us to find a moderate excess ($\sim 2.4- 3\sigma$ depending on the assumption of the neutron-distribution radius) beyond the known backgrounds including neutrino-induced ones and beam-related ones. 
We then find a set of model parameters to accommodate the excessive number of events, keeping in mind the aforementioned dark matter scenario as our underlying model assumption.  
We also study the expected dark matter signal sensitivities at our benchmark detectors of the COHERENT, JSNS$^2$, and CCM experiments, considering various dark sector gauge boson models. 
Our results suggest that their experimental reaches in the associated model parameter space can be beyond the existing limits.
For all these experiments, we discuss the appearance of dark matter (in the detector) which makes them very robust compared to the beam-dump searches where the disappearance of dark photon is used to investigate the dark matter parameter space.

The paper is organized as follows. 
We begin by describing dark matter scenarios we consider, in terms of production and detection of dark matter at neutrino experiments, in Section~\ref{sec:models}.
We then give a brief review on our benchmark experiments which can possess decent sensitivities to our dark matter signal in Section~\ref{sec:exp}.
We performed simulation with the \texttt{GEANT}4 package to assess dark matter production in each experiment channel-by-channel, and our results are reported in Section~\ref{sec:flux}.  
A brief discussion on the timing spectrum of dark matter events and its implication on our search strategy are given in Section~\ref{sec:timing}. 
Section~\ref{sec:analysis} describes the general search strategy we propose.
In Section~\ref{sec:interpretations}, we discuss how to interpret the results from data analyses. We first show our example data analyses for the existing CsI and LAr data of the COHERENT experiment in Sections~\ref{sec:explain} and \ref{sec:explain_lar}, respectively, and then focus on the signal sensitivities expected at our benchmark detectors for example dark matter scenarios in Section~\ref{sec:constrain}.
Section~\ref{sec:recast} is devoted to discussing how to reinterpret our results in the context of other dark matter scenarios.
Our conclusions and outlook appear in Section~\ref{sec:conclusions}.
Finally, calculation of phase-space suppression factors of cascade photons and derivation of the timing spectrum for an example scenario are provided in appendices \ref{sec:appA} and \ref{sec:appB}, respectively.

\section{Models: Dark Matter Scenarios \label{sec:models}}

In this section, we discuss various new physics scenarios which can be probed with the proposed search strategy.
The most topologically-minimal possibility is the case where a certain (new physics) intermediary state, say $A$, is produced and disintegrates into lighter particles some of which are stable enough to reach a detector and leave some visible signature, for example, through scattering off a nucleus or an electron in the detector:
\begin{equation}
    A \to \chi + {\rm others},~~~\chi+e^-/N \to \chi^{(\prime)} + e^{-}/N^{(\prime)}\,.
\end{equation}
Here $\chi$, $e^-$ and $N$ denote stable particle species, electron and nucleus, respectively, and $\chi'$ represents the possibility that $\chi$ turns into a different species in the scattering process~\cite{TuckerSmith:2001hy,Izaguirre:2014dua,Kim:2016zjx,Giudice:2017zke}.
More generally, such an intermediary particle may undergo a multi-step cascade decay depending on the model details, and some of the stable decay products (including neutrinos) again may repeat the aforementioned procedure:
\begin{equation}
    A \to B \to \cdots \to C \to \chi + {\rm others},~~~\chi+e^-/N \to \chi^{(\prime)} + e^{-}/N^{(\prime)}\,,
\end{equation}
where $B$ and $C$ denote other intermediary states and where decay products in each decay step are omitted for simplicity. 
Some of the decay products may be SM particles so that they directly leave visible signatures at the detector. 
While such a possibility is interesting {\it per se} and similar strategies are applicable, we focus on the case with stable new physics particles such as dark matter in this study.

\subsection{Production of dark matter \label{sec:production}}

Among possible new physics scenarios, we consider the production of dark matter $\chi$ by the decay of a dark gauge boson $X$.
The relevant interaction Lagrangian is given by
\begin{equation}
    \mathcal{L}_{X,{\rm prod}} \supset \sum_f  \kappa_f^X x_f^X X_\mu \bar{f} \gamma^\mu f + \kappa_D^X X_\mu \bar{\chi} \gamma^\mu \chi\,, \label{eq:lagprod}
\end{equation}
where $x_f^X$ is the gauge charge of SM fermion species $f$ and $\kappa_f^X$ denotes the coupling constant associated with the dark gauge boson $X$.  
By contrast, $\kappa_D^X$ parameterizes the dark sector coupling of $X$ to $\chi$. 
We emphasize that our approach is generic and various types of gauge bosons can take the role of $X$. Possible examples of $X$ include the following models.
\begin{itemize}
    \item Dark photon: 
    A dark sector photon is connected to the SM sector photon via a kinetic mixing~\cite{Okun:1982xi,Galison:1983pa,Holdom:1985ag}.
    Phenomenology of dark photon coupled to dark matter has been extensively studied in an ample amount of literature, e.g.,~\cite{Huh:2007zw,Pospelov:2007mp,Chun:2010ve}.
    \item Baryo-philic U(1)$_B$: 
    This new U(1) boson is assumed to couple to baryon number, hence exclusively to quarks (at the tree level)~\cite{Nelson:1989fx,Rajpoot:1989jb}.
    Possible are models where dark matter is charged under a U(1)$_B$ symmetry~\cite{Batell:2014yra,deNiverville:2015mwa}.
    \item U(1)$_{B-L}$: 
    This gauge boson couples to the difference between the baryon and the lepton numbers and is often invoked in some grand unified theory models~\cite{Davidson:1978pm,Wilczek:1979hc}.
    Various phenomenological studies in connection with dark matter have been performed, e.g., asymmetric dark matter~\cite{Kaplan:2009ag}.
    \item Lepto-philic U(1)$_L$ and U(1)$_{L_i - L_j}$: 
    Unlike the previous models, the relevant gauge bosons exclusively couple to leptons (at the tree level)~\cite{Rajpoot:1989jb, He:1990pn}.
    Dark matter can be coupled to these gauge bosons, resulting in interesting phenomenology~\cite{Bi:2009uj,Kim:2015fpa, Foldenauer:2018zrz}.
    \item U(1)$_{T3R}$: 
    This is a low energy  SU(3)$_C\times$SU(2)$_L\times$U(1)$_Y\times$U(1)$_{T3R}$ anomaly-free model where the U(1)$_{T3R}$ gets broken at sub-GeV scale~\cite{Dutta:2019fxn}.
    The model contains a sub-GeV scalar, a sub-GeV gauge boson, and a dark matter particle.
    Two scenarios from~\cite{Dutta:2019fxn} of this model are interesting for this paper which we describe as model 1: the new symmetry involves first-generation right-handed quarks and right-handed electron and model 2: the new symmetry involves first-generation right-handed quarks and right-handed muon.
\end{itemize}
Table~\ref{tab:modelsummary} summarizes these models relevant for the benchmark experiments, which we will discuss in Section~\ref{sec:exp}, and their respective coupling constants and tree-level gauge charges for the SM fermions.
For example, if the model of interest is dark photon, $X$ is replaced by $A'$ and the coupling constant $\kappa_f^X$ reads $\epsilon e$ with $\epsilon$ being the kinetic mixing parameter between the SM U(1)$_{\rm EM}$ gauge boson and the dark sector U(1) gauge boson.
For U(1)$_{T3R}$, the charge assignment for the particles  of model 2 is mentioned in the parentheses only when they differ compared to model 1.
Note that the U(1)$_B$ model and the U(1)$_L$ model are anomalous, while the others can be anomaly-free. 

\begin{table}[t]
    \centering
    \resizebox{\columnwidth}{!}{%
    \begin{tabular}{c|c c c c c c c}
    \hline \hline
    Model & $A'$ & $B$ & $L$ & $B-L$ & $L_e - L_\mu$ & $L_e-L_\tau$&$T_{3R}$ [model 1(2)]\\
    \hline
    $\kappa_f^X$  & $\epsilon e$  & $g_B$ & $g_L$ & $g_{B-L}$ & $g_{L_e - L_\mu}$ & $g_{L_e-L_\tau}$ & $g_{T_{3R}}$ \\
    $x_{u,c,t}^X$ & 2/3& 1/3 & 0 & 1/3 & 0 & 0&-2, 0, 0\\
    $x_{d,s,b}^X$ & $-1/3$& 1/3 & 0 & 1/3 & 0 & 0&2, 0, 0 \\
    $x_{e,\mu,\tau}^X$ & $-1$ &  0 & 1 & $-1$ & 1, $-1$, 0 & 1, 0, $-1$ & 2 (0), 0 (2), 0\\
    $x_{\nu_e,\nu_\mu,\nu_\tau}^X$ & 0 & 0 & 1 & $-1$ & 1, $-1$, 0 & 1, 0, $-1$&0, 0, 0 \\
    \hline 
    Production & P: 1, 2, 3, 4 & P: 1, 2, 3 & P4 & P: 1, 2, 3, 4 & P4 & P4 & P: 1, 2, 3, 4 (1, 2, 3) \\
    Detection & D1, D2 & D1 & D2 & D1, D2 & D2 & D2&D1, D2 (D1) \\
    \hline \hline
    \end{tabular}%
    }
    \caption{Possible models relevant to the benchmark experiments enumerated in Section~\ref{sec:exp} and their respective coupling constants $\kappa_f^X$ and tree-level gauge charges $x_f^X$ for the SM fermions.
    The (tree-level) production channels and the detection channels are summarized in the last two rows.}
    \label{tab:modelsummary}
\end{table}

The first term in Eq.~\eqref{eq:lagprod} essentially allows for production of dark gauge boson. 
As mentioned in the introductory section, four different production mechanisms of dark gauge boson can be mainly relevant.
\begin{itemize}
    \item {\bf Meson decays} (P1): 
    Neutral mesons such as $\pi^0$ and $\eta$ created in nuclear reactions decay to a pair of SM photons where one of the photons may be replaced by a dark gauge boson $X$. 
    To distinguish $\pi^0$ in this category from the one in the charge exchange processes (explained in the third item), we henceforth call it primary $\pi^0$ while the other one is labeled as secondary $\pi^0$.
    Unlike the secondary $\pi^0$, the primary $\pi^0$ is boosted toward the forward direction.
    So, the detectors located in the backward direction do not benefit from this production mechanism much, if the incoming beam is highly energetic.
    \item {\bf $\pi^-$ absorption process} (P2): 
    $\pi^-$ created in the proton/electron dump experiment can be absorbed to a nucleus by the so-called Panofsky process $\pi^-+p \to  n + \gamma$~\cite{Panofsky:1950he} where the SM photon may be replaced by a dark gauge boson $X$. 
    This process is efficient once $\pi^-$ becomes non-relativistic.
    Therefore, the mesic state formed by a $\pi^-$ and a proton is produced nearly at rest, and $X$ is emitted isotropically.
    In a complex atom, the monochromatic Panofsky photon gets through nuclear reactions with nearby nucleons and electrons, resulting in a bunch of soft photons, neutrons, and possibly with an element different from the original target one.
    However, once a dark gauge boson is emitted, it hardly interacts with nearby nucleons and electrons unless the associated coupling is sizable enough.  
    Therefore, it is expected that the energy of the dark gauge boson is single-valued.
    \item {\bf Charge exchange processes} (P3): 
    $\pi^{-(+)} + p(n) \to n(p) + \pi^0$, followed by the decay of $\pi^0$ to an ordinary photon and an $X$.
    These processes are efficient once $\pi^\pm$ becomes non-relativistic, just like the absorption process.
    Therefore, $\pi^0$ is emitted isotropically, and so is $X$ coming from the $\pi^0$ decay explained in P1.
    \item {\bf $e^\pm$-induced cascade} (P4):\footnote{A work~\cite{Celentano:2020vtu} to perform a dedicated study on the cascade photon contribution appears on the same day.}
    Primary particles produced in the target by beam collision lose their energy by ionization, creating electrons which subsequently undergo electromagnetic cascade showering.
    Therefore, expected are a copious number of cascade photons some of which may be replaced by an $X$ which charges the electron non-trivially.
\end{itemize}
Depending on the underlying model assumption, all or part of the above-listed production channels are relevant. 
For example, a dark photon $A'$ can be produced by P1 through P4 (see Table~\ref{tab:modelsummary} for relevant production channels of each model). 
The dark matter fluxes from these contributions obviously depend on the beam energy. 
We shall discuss and compare relative production rates, which are evaluated by the \texttt{GEANT4} simulation code package, in the context of our benchmark experiments in Section~\ref{sec:exp}.

The second term in Eq.~\eqref{eq:lagprod} governs the decay of $X$ to a dark matter pair, as far as the mass of dark gauge boson is greater than twice the mass of dark matter. 
The associated decay width $\Gamma_{X\to 2\chi}$ is expressed as
\begin{equation}
    \Gamma_{X\to 2\chi} = \frac{1}{12\pi}(\kappa_D^X)^2  m_{X}\left(1+\frac{m_\chi^2}{m_{X}^2} \right) \sqrt{1-\frac{4 m_\chi^2}{m_{X}^2}}\,,
\end{equation}
where $m_{X}$ and $m_\chi$ stand for the masses of dark gauge boson and dark matter, respectively.
If $X$ is heavy enough, it is allowed to decay directly to SM particles as far as the associated mass hierarchy is kinematically allowed. 
Therefore, the branching fraction to a dark matter pair, BR$_{X\to 2\chi}$, is determined together with all allowed SM decay modes. 
Since we are interested in the case where $X$ predominantly decays to a dark matter pair, i.e., BR$_{X\to 2\chi}\approx 1$, at least one of the following conditions should be satisfied:
($i$) $\kappa_D^X \gg \kappa_f^X x_f^X $, ($ii$) $m_X < 2m_f$, and ($iii$) $x_\ell \ll 1$ and $m_X < 2 m_\pi$ with $\ell$ being SM leptons.
Note that the first two cases are relevant for both baryo-philic and lepto-philic models, while the condition ($iii$) is relevant for baryo-philic models.

\subsection{Detection of dark matter \label{sec:detectiondm}}

Once dark matter $\chi$ reaches a detector, it can leave a scattering signature through interactions with detector material. 
We point out that a different mediator, say $V$, may take care of the scattering part. Therefore, the relevant interaction Lagrangian contains exactly the same type of operators as in Eq.~\eqref{eq:lagprod},
\begin{equation}
    \mathcal{L}_{V,{\rm scatter}} \supset \sum_f  \kappa_f^V x_f^V V_\mu \bar{f} \gamma^\mu f + \kappa_D^V V_\mu \bar{\chi} \gamma^\mu \chi\,, \label{eq:vscat}
\end{equation}
with $X_\mu$, $\kappa_f^X$, $x_f^X$, and $\kappa_D^X$ replaced by $V_\mu$, $\kappa_f^V$, $x_f^V$, and $\kappa_D^V$, respectively.   
Obviously, in the minimal scenario where a single mediator governs scattering as well as production, $V_\mu$, $\kappa_f^V$, $x_f^V$, and $\kappa_D^V$ are simply identified as $X_\mu$, $\kappa_f^X$, $x_f^X$, and $\kappa_D^X$, and vice versa. 
We now consider two possible scenarios, the scattering of dark matter off either a nucleus or an electron since the example dark gauge bosons in Table~\ref{tab:modelsummary} can interact with quarks and/or electrons.
\begin{itemize}
\item {\bf Nucleus scattering} (D1): 
Given the energy and momentum, $E_\chi$ and $p_\chi$, of incoming dark matter, the differential scattering cross section in recoil energy $E_{r,N}$ of the target nucleus is expressed as~\cite{Dutta:2019nbn}
\begin{equation}
    \frac{d\sigma}{dE_{r,N}}=\frac{(\kappa_f^V \kappa_D^V)^2 (Q_{\rm eff}^V)^2 \cdot |F_V|^2}{4\pi p_\chi^2(2m_NE_{r,N}+m_V^2)^2} \left\{2E_\chi^2 m_N\left( 1-\frac{E_{r,N}}{E_\chi}-\frac{m_NE_{r,N}}{2E_\chi^2}\right) +m_N E_{r,N}^2\right\},
\label{eq:nuclear_xs}
\end{equation}
where $m_N$ is the mass of the target nucleus and where $F_V$, which is a function over $m_N$ and $E_{r,N}$, denotes the form factor associated with the dark gauge boson $X$. 
Here $Q_{\rm eff}^V$ is an effective prefactor for a given nucleus which is atomic number $Z$ for the $A'$ model, atomic mass number $A$ for the $B$ and $B-L$ models, $2(A-2Z)$ for the $T_{3R}$ model, and a small loop-induced factor for the other models. 
We note that although the model here considers the dark matter flavor-conserving interaction only [i.e., the second term in Eq.~\eqref{eq:vscat}] for simplicity, one can include the term responsible for the scenario where $\chi$ upscatters to a heavier (or excited) unstable state, say $\chi^*$ via an exchange of $V$~\cite{Izaguirre:2014dua,Kim:2016zjx,Izaguirre:2017bqb,Giudice:2017zke,Chatterjee:2018mej,Berlin:2018pwi,Heurtier:2019rkz,Kim:2020ipj,DeRoeck:2020ntj}.
We reserve this possibility for future work.
\item {\bf Electron scattering} (D2):
If the dark gauge boson under consideration allows for the interaction of dark matter with electrons in the detector material, the dark matter can manifest itself as an electron recoil. 
The associated differential scattering cross section in recoil energy $E_{r,e}$ is given by~\cite{Agashe:2014yua,Kim:2016zjx}
\begin{eqnarray}
    \frac{d\sigma}{dE_{r,e}}&=&\frac{Z(x_f^V\kappa_f^V \kappa_D^V)^2 m_e^2}{\pi \lambda(s,m_e^2,m_\chi^2)\left\{2m_e(m_e-E_{r,e})-m_V^2 \right\}^2} \nonumber \\
    &\times& \left[m_e \left\{E_\chi^2+(m_e+E_\chi-E_{r,e})^2 \right\} +\left(m_e^2 + m_\chi^2\right) \left(m_e - E_{r,e} \right) \right],
\label{eq:electron_xs}
\end{eqnarray}
where $s=m_e^2+2E_\chi m_e+m_\chi^2$ and $\lambda$ is a kinematic triangular function defined as $\lambda(x,y,z)\equiv (x-y-z)^2-4yz$.
Here an overall factor $Z$ takes into account the number of electrons in a target atom.
As in the case of nucleus scattering, one can consider the scenario where $\chi$ scatters off an electron to $\chi^*$ via an exchange of $V$.
\end{itemize}
The detection channels available are determined by the underlying model assumption.
We summarize relevant detection channels of each model in Table~\ref{tab:modelsummary} with $X$ replaced by $V$.

It is noteworthy that the typical recoil energy in the nucleus scattering is much smaller than that in the electron scattering.
The reason is because the typical mass scale of dark matter under consideration is much smaller than the mass of the target nucleus while being larger than or comparable to the mass of the target electron.
Therefore, a sufficiently small energy threshold ($E_r^{\rm th}$) is demanded in order to observe nucleus recoil signals. 
In the next section, we will see that the detectors in COHERENT and CCM are designed to be sensitive enough to the nucleus recoil, whereas JSNS$^2$ possesses a good sensitivity to the electron recoil due to a relatively larger energy threshold of its detector.

\section{Benchmark Experiments and Simulations \label{sec:experiments}}

As briefly discussed in the Introduction, the key feature behind the proposed search strategy is to veto the prompt neutrinos and the delayed neutrinos by an energy cut and a timing cut, respectively. 
Indeed, the effectiveness of the search strategy can be maximized when the (dominant) neutrino sources, i.e., $\pi^\pm$ and $\mu^\pm$, decay to neutrino(s) (almost) at rest. 
First of all, under such circumstances, prompt neutrinos are nearly single-valued in energy ($\sim30$ MeV) due to the two-body decay of $\pi^\pm$, so the resultant energy deposit is upper-bounded. 
This implies that the prompt neutrino events can be significantly suppressed by imposing an energy cut associated with such an upper bound. 
However, if $\pi^\pm$ were substantially boosted, energies of prompt neutrinos would be distributed and extended toward the higher energy regime. 
Therefore, one would need to impose too hard an energy cut for achieving the similar level of background rejection, which would result in significant signal rejection.  
Second, when it comes to the muon-induced neutrinos, the relatively long lifetime of muon delays its decay products reaching the detector. 
Therefore, an upper-bounded timing cut can substantially remove delayed neutrino events, provided that the muons are not so boosted that they dominate in the prompt timing bins.
Finally, if dark gauge bosons were significantly boosted forward, the dark matter signal flux could be forward-directed. 
As a result, detectors would have to be located in the forward region where other backgrounds such as beam-related neutrons would be increasingly considerable.

\subsection{Benchmark experiments \label{sec:exp}} 

Taking all these aspects into consideration, experiments with a low-energy beam source are more relevant to the proposed signal search strategy. 
Of possible experiments, we take COHERENT, JSNS$^2$, and CCM as our benchmark cases. 
We first give a brief review on these experiments, summarizing their key specifications in Table~\ref{tab:expspec} for convenience.  

\begin{table}[t]
    \centering
    \resizebox{\columnwidth}{!}{%
    \begin{tabular}{c|c c c c}
    \hline \hline
    \multirow{2}{*}{Experiment}     & $E_{\rm beam}$ & POT & \multirow{2}{*}{Target} & \multirow{2}{*}{Detector: mass, distance, angle, $E_r^{\rm th}$} \\
     & [GeV] & [yr$^{-1}$] & &  \\
   \hline
   COHERENT & \multirow{2}{*}{1}  & \multirow{2}{*}{$1.5\times 10^{23}$} & \multirow{2}{*}{Hg} & CsI[Na]: 14.6~kg, 19.3~m, 90$^\circ$, 6.5~keV \\
   \cite{Akimov:2017ade, Akimov:2018ghi, Akimov:2019xdj} & & & & LAr: 24~kg (0.61~ton), 28.4~m, 137$^\circ$, 20~keV \\
    JSNS$^2$~\cite{Ajimura:2015yux, Ajimura:2017fld, privateJSNS2} & 3 & $3.8\times 10^{22}$ & Hg & Gd-LS: 17~ton, 24~m, 29$^\circ$, 2.6~MeV\\
    CCM~\cite{CCM, dunton_2019, privateCCM} & 0.8 & $1.0\times 10^{22}$ & W & LAr: 7~ton, 20~m, 90$^\circ$, 25~keV\\
    \hline \hline
    \end{tabular}%
    }
    \caption{Key specifications of benchmark experiments and detectors under consideration.
    All three experiments use a proton beam, and the POT values are expected spills for 5,000 hours operation per year.
    The mass of the liquid argon detector in parentheses in COHERENT is for a future upgrade.
    }
    \label{tab:expspec}
\end{table}

\begin{itemize}
\item {\bf COHERENT}~\cite{Akimov:2017ade, Akimov:2018ghi, Akimov:2019xdj}: 
The main mission of the COHERENT experiment is to make the first direct measurement of Coherent Elastic Neutrino-Nucleus Scattering predicted by the Standard Model.
The 1.4~MW beam of 1 GeV protons\footnote{This proton beam could be upgraded to 2.4 MW and 1.3 GeV with a planned second target station~\cite{osti_1185891}.} (0.6~$\mu$s wide pulses at a rate of 60 Hz) at Oak Ridge National Laboratory impinges on a mercury target by the rate of $\sim8.8\times 10^{15}$ POT per second, which provides $\sim1.5\times 10^{23}$ POT during operation of 5,000 hours per year. 
The expected neutrino flux at 20~m is $4.3\times 10^7~{\rm cm}^{-2}{\rm s}^{-1}$.
There are six different detectors in the Neutrino Alley located $\sim 5$~m below from the target ($\sim8$ meter water-equivalent overburden). 
Of them, we consider the 14.6~kg CsI[Na] detector located at a distance of 19.3~m from the target and LAr detectors with 24~kg (called CENNS-10) and future-upgraded 0.61~tons of fiducial volume being 28.4~m away from the target.

\item {\bf JSNS$^2$}~\cite{Ajimura:2015yux, Ajimura:2017fld, privateJSNS2}:
The J-PARC Sterile Neutrino Search at the J-PARC Spallation Neutron Source (JSNS$^2$) experiment, which has started data taking in June 2020, aims to probe the existence of neutrino oscillations with $\Delta m^2$ around $1~{\rm eV}^2$. 
The 1~MW beam of 3 GeV protons (two 0.1~$\mu$s wide pulses separated by 0.44~$\mu$s with a repetition rate of 25 Hz) is incident on a mercury target by the rate of $\sim 2.1\times 10^{15}$ POT per second corresponding to $\sim 3.8\times 10^{22}$ POT for 5,000 hours operation per year.
The produced neutrinos reach a gadolinium(Gd)-loaded liquid-scintillator (LS) detector which has a fiducial mass of 17 tons and is placed at a distance of 24~m from the target. 
The Gd-loaded LS detector is surrounded by $\sim30$ tons of unloaded LS which vetoes the background signals coming from outside.
The expected $\bar{\nu}_\mu$ neutrino flux at this detector is $1.8\times 10^{14}~{\rm cm}^{-2}{\rm year}^{-1}$ ($\approx 5.7\times 10^6~{\rm cm}^{-2}{\rm s}^{-1}$).

\item {\bf CCM}~\cite{CCM, dunton_2019, privateCCM}:
Coherent Captain-Mills (CCM) detector was proposed to study coherent neutrino scattering.
The $\sim80$ kW beam of 800 MeV protons (0.29~$\mu$s wide triangular pulses at a rate of 20 Hz) is bombarded on a tungsten target.
The detector is made of a total (fiducial) mass of 10 tons (7 tons) LAr and located 20~m away from the Lujan target at Los Alamos National Laboratory. 
$\sim5.6\times 10^{14}$ POT/second (equivalent to $\sim 1.0\times 10^{22}$ POT for 5,000 hours operation per year) will yield a neutrino flux of $4.7\times 10^5~{\rm cm}^{-2}{\rm s}^{-1}$ at the detector for each neutrino species. 
\end{itemize}

\subsection{Estimating dark matter fluxes \label{sec:flux}}

As obvious from the production mechanisms of dark matter that were discussed in Section~\ref{sec:models}, precise knowledge on production rates of relevant mesons such as $\pi^\pm$, $\pi^0$, and $\eta$ is crucial for better estimates on (differential) dark matter flux. 
To this end, we have performed simulations using the \texttt{GEANT}4 10.5 with the FTFP\_BERT library ~\cite{Agostinelli:2002hh}.
We take the target specification of JSNS$^2$ from Ref.~\cite{Ajimura:2017fld} without considering the (nearly irrelevant) modules around the targets. 
The specifications for COHERENT and CCM are not publicly available, so we adopt that of JSNS$^2$ as a substitute. 
However, we find that the numbers reported here do not depend on the target details much~\cite{private}, so we simply quote them in our data analysis. 

\begin{table}[t]
    \centering
    \begin{tabular}{c|c|c|c|c|c|c}
    \hline \hline 
    Experiment     & \multicolumn{2}{c}{COHERENT} & \multicolumn{2}{|c|}{JSNS$^2$} & \multicolumn{2}{c}{CCM}  \\
    \hline 
         & $\pi^+$ & $\pi^-$ & $\pi^+$ & $\pi^-$ & $\pi^+$ & $\pi^-$ \\
    \hline \hline
    $N_{\pi^\pm}$ & 0.1098 & 0.0470 & 0.5260 & 0.4962 & 0.0665 & 0.0259 \\
    \hline 
    Decay ($\pi \to \mu+\nu$) & 0.0803 & 0.0001 & 0.2603 & 0.0019 & 0.0520 & 0.00004 \\
    Inelastic (w. $\pi^0$) & 0.0016 & 0.0004 & 0.0214 & 0.0124 & 0.0006 & 0.0002 \\
    Inelastic (w.o. $\pi^0$) & 0.0239 & 0.0113 & 0.2081 & 0.2071 & 0.0112 & 0.0053 \\
    Capture at rest (w. $\pi^0$) & 0.0 & 0.0 & 0.0 & 0.0 & 0.0 & 0.0 \\
    Capture at rest (w.o. $\pi^0$) & 0.0 & 0.0333 & 0.0 & 0.2443 & 0.0 & 0.0192 \\
    Transportation & 0.0037 & 0.0017 & 0.0351 & 0.0296 & 0.0022 & 0.0009 \\
    \hline \hline
    $N_{\pi^0}$ & \multicolumn{2}{c}{0.1048} & \multicolumn{2}{|c|}{0.6142} & \multicolumn{2}{c}{0.0633}  \\
    $N_{\eta}$ & \multicolumn{2}{c}{0.0} & \multicolumn{2}{|c|}{0.0015} & \multicolumn{2}{c}{0.0}  \\
    $N_{K^+}$ & \multicolumn{2}{c}{0.0} & \multicolumn{2}{|c|}{0.0061} & \multicolumn{2}{c}{0.0}  \\
    $N_{K^-}$ & \multicolumn{2}{c}{0.0} & \multicolumn{2}{|c|}{0.0001} & \multicolumn{2}{c}{0.0}  \\
    \hline \hline
    \end{tabular}
    \caption{A summary of our \texttt{GEANT}4 10.5 (FTFP\_BERT) simulation results with $10^5$ protons struck on the target of each benchmark experiment. 
    $N_i$ denotes the fractional number of particles of species $i$ per POT. 
    The 4th through 9th rows describe further processes that produced charged pions undergo.
    See the text for details. }
    \label{tab:sim}
\end{table}

Our simulation results with $10^5$ protons struck on the target of each benchmark experiment are summarized in Table~\ref{tab:sim}. 
We first report the fractional number of charged pions per proton in the third row $N_{\pi^\pm}$. 
The next six rows describe further processes that the produced $\pi^\pm$ undergo, so the summation of the numbers therein returns $N_{\pi^\pm}$. 
As the JSNS$^2$ Collaboration reported the numbers corresponding to those in the third and fourth rows~\cite{Ajimura:2017fld}, we are able to compare them and find that our simulation results are in good agreement with their numbers. 
We also compared our simulation with COHERENT simulation data~\cite{private}, and found that ours reproduces their numbers. 
``Decay'' is for the case where a produced pion decays to a muon and a neutrino whether or not it decays in flight. 
The two rows of ``Inelastic'' are for the cases where moving pions disappear, creating a neutral pion in the final state or not. 
By contrast, the next two rows consider the cases where charged pions are captured at rest, creating a neutral pion or not. 
Finally, ``Transportation'' describes the number of $\pi^\pm$ which simply escape from the target.
The last four rows report the fractional numbers of other mesons such as $\pi^0$, $\eta$, and $K^\pm$. 
Be aware that $N_{\pi^0}$ do not include the $\pi^0$ induced by $\pi^\pm$, i.e., only primary $\pi^0$ and $\pi^0$ from heavier-meson decays are taken into account. 

Several observations are made in order. 
First of all, only a tiny fraction of positively charged pions create neutral pions while the rest of them either decay to muon or disappear without involving neutral pion.
Therefore, their contribution to dark matter production is subdominant. 
Second, most of the negatively charged pions get through either ``Inelastic (w.o. $\pi^0$)'' or ``Capture at rest (w.o. $\pi^0$)'', i.e., the absorption process involving a mesic state is dominant. 
For the ``Inelastic'' case, our simulation study suggests that the mesic state be not much boosted, and therefore, it is still a good approximation that its decay is isotropic in the laboratory frame.
Finally, other mesons heavier than pion could be considered as sources of dark gauge boson. 
JSNS$^2$ and COHERENT may create a small fraction of $\eta$, $\mathcal{O}(10^{-3})$ $\eta$ per POT.\footnote{Production of the $\eta$ meson at COHERENT occurs by nuclear motion~\cite{Cassing:1991jr, Akimov:2019xdj}, which is not yet implemented in the \texttt{GEANT4} package.
We include it in our analysis in addition to other particles that can be simulated by \texttt{GEANT4}.}
JSNS$^2$ also contains $\sim0.006$ $K^+$ per POT and $\mathcal{O}(10^{-4})$ $K^-$ per POT, because the electric charge consideration of the proton-nucleus collision makes it easier to have $K^+$ than $K^-$.
Nevertheless, the $K^+$ contribution is negligible.
However, for CCM, we have not found any such particles in our simulation data because they are relatively too heavy to be produced, given the allowed energy budget.\footnote{One caveat to keep in mind is that production of these heavier mesons with low-energy beam sources depends on the nuclear models to consider~\cite{Akimov:2019xdj}. 
One would obtain them by choosing appropriate physics options available in the \texttt{GEANT}4 package.}

\begin{figure}[t]
    \centering
    \includegraphics[width=7.4cm]{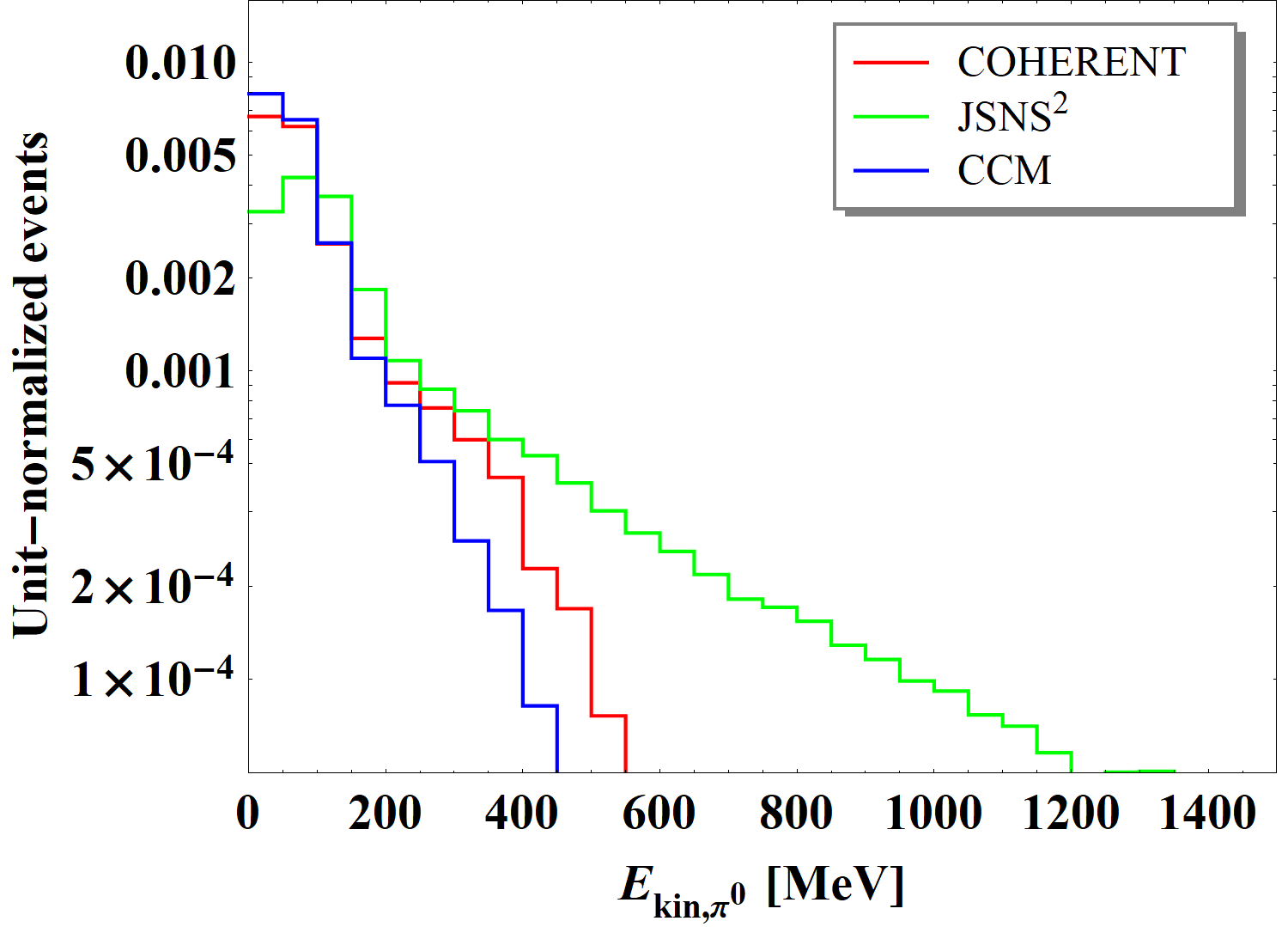} \hspace{0.2cm}
    \includegraphics[width=7.05cm]{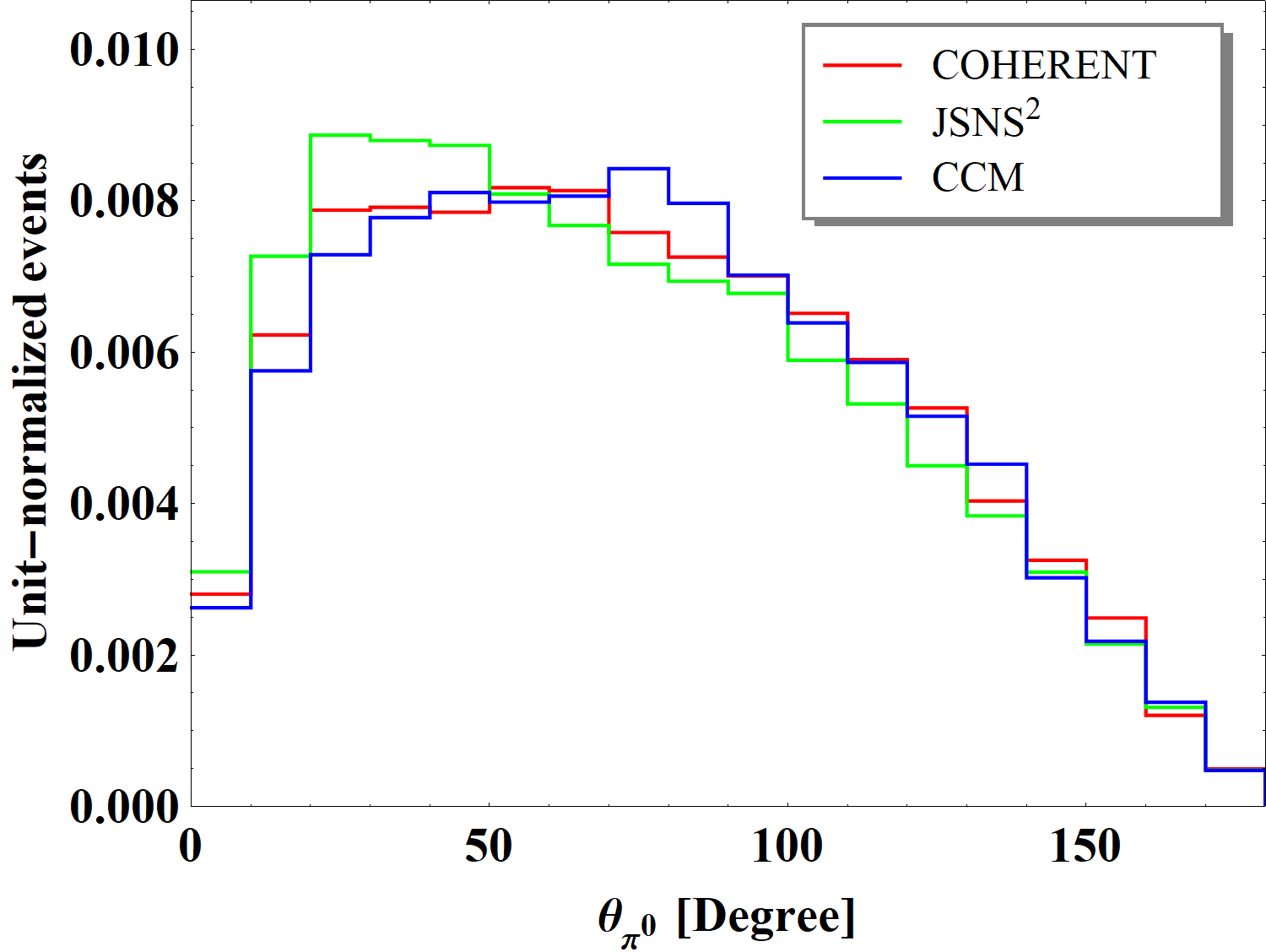}
    \caption{Unit-normalized $\pi^0$ kinetic energy spectra (left) and angular spectra (right) in the COHERENT (red), JSNS$^2$ (green), and CCM (blue) experiments. 
    We use \texttt{GEANT}4 10.5 with the FTFP\_BERT library~\cite{Agostinelli:2002hh} for our simulation, generating $10^5$ protons incident on the targets. 
    The angle is measured from the incident proton beam line.}
    \label{fig:pionspec}
\end{figure}

When it comes to $\pi^0$, its spectral behaviors in energy and flying direction are important to estimate the dark matter flux. 
In general, $\pi^0$ is not produced at rest, so the resulting $\pi^0$ flux lies slightly in the forward direction. 
Therefore, the $\pi^0$-induced dark matter flux should be carefully estimated according to the detector location relative to the beam line and the target.
Figure~\ref{fig:pionspec} shows unit-normalized $\pi^0$ kinetic energy spectra (left) and angular spectra (right) in our benchmark experiments, COHERENT (red), JSNS$^2$ (green), and CCM (blue).
Here the angle variable $\theta_{\pi^0}$ is measured from the incident proton beam line. 
Looking at the spectra of $E_{{\rm kin},\pi^0}$, we see that neutral pions carry non-negligible energy although peaks are around 50~MeV. 
Clearly, $\pi^0$s of JSNS$^2$ are more inclined to be boosted as the associated proton beam delivers more energy compared to the other two. 
Therefore, the dark matter particles produced in JSNS$^2$ can be as energetic as $\sim 1$~GeV. 
This implies that $\pi^0$ in JSNS$^2$ are more forward-directed with respect to the beam line. 
The right panel of Figure~\ref{fig:pionspec} confirms this expectation as the $\pi^0$ flux peaks around 25 degrees. 
On the other hand, the $\pi^0$ fluxes of COHERENT and CCM are more or less flat in-between 25 and 100 degrees. 
Thus, a decent level of $\pi^0$ flux can be directed to some of backward-located detectors. 

\begin{figure}[t]
    \centering
    \includegraphics[width=7.4cm]{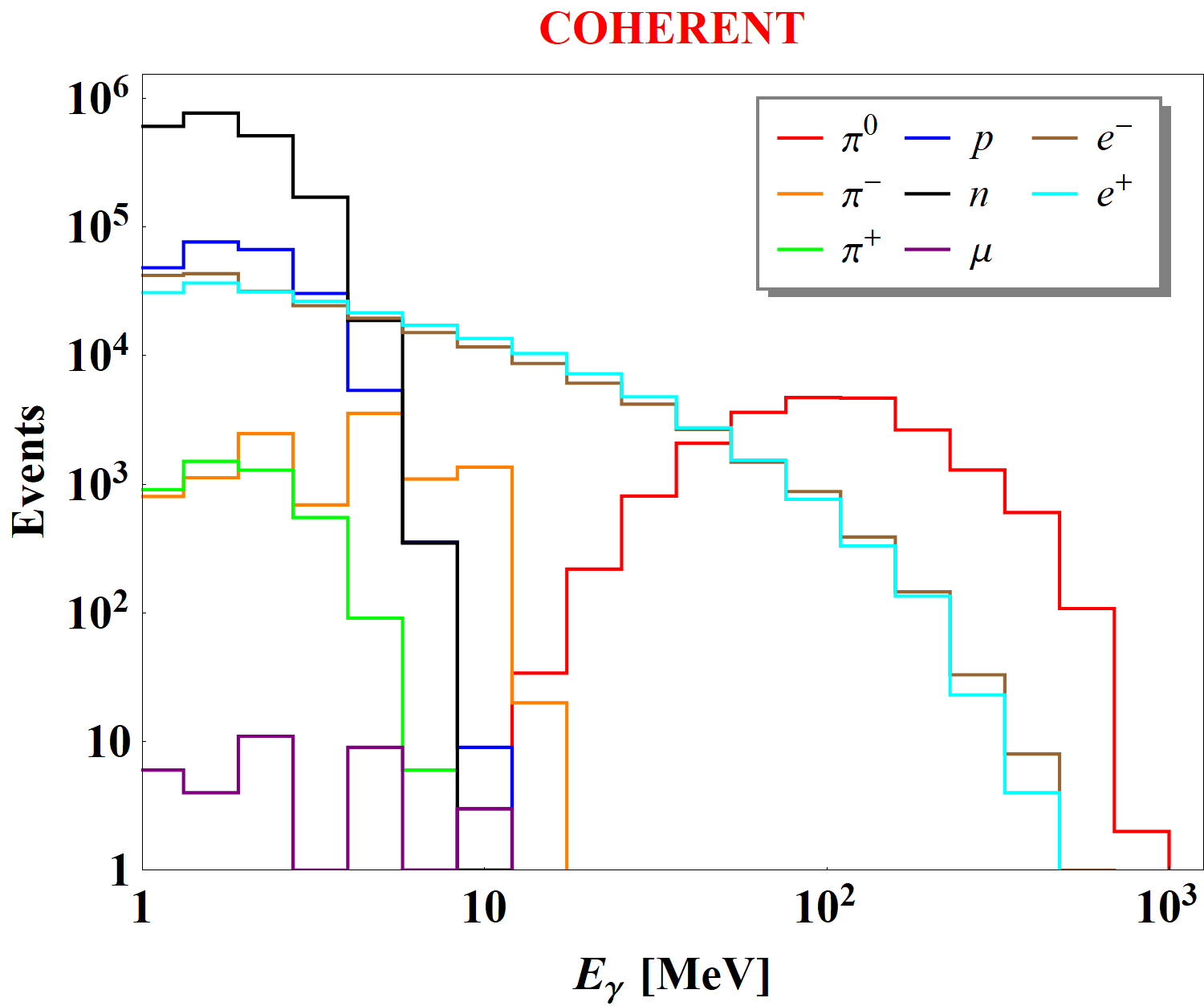}
    \includegraphics[width=7.4cm]{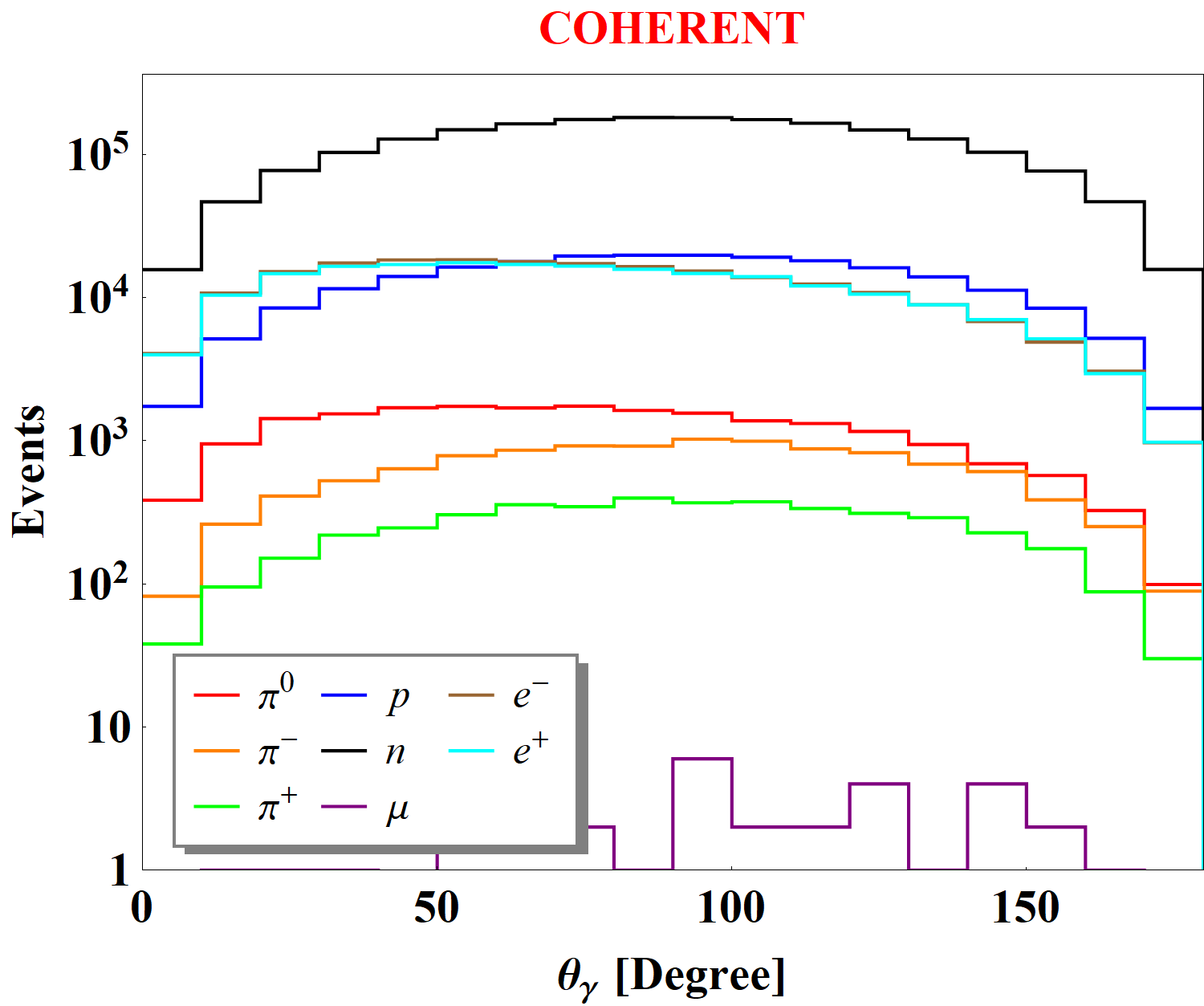} \\
    \vspace{0.3cm}
    \includegraphics[width=7.4cm]{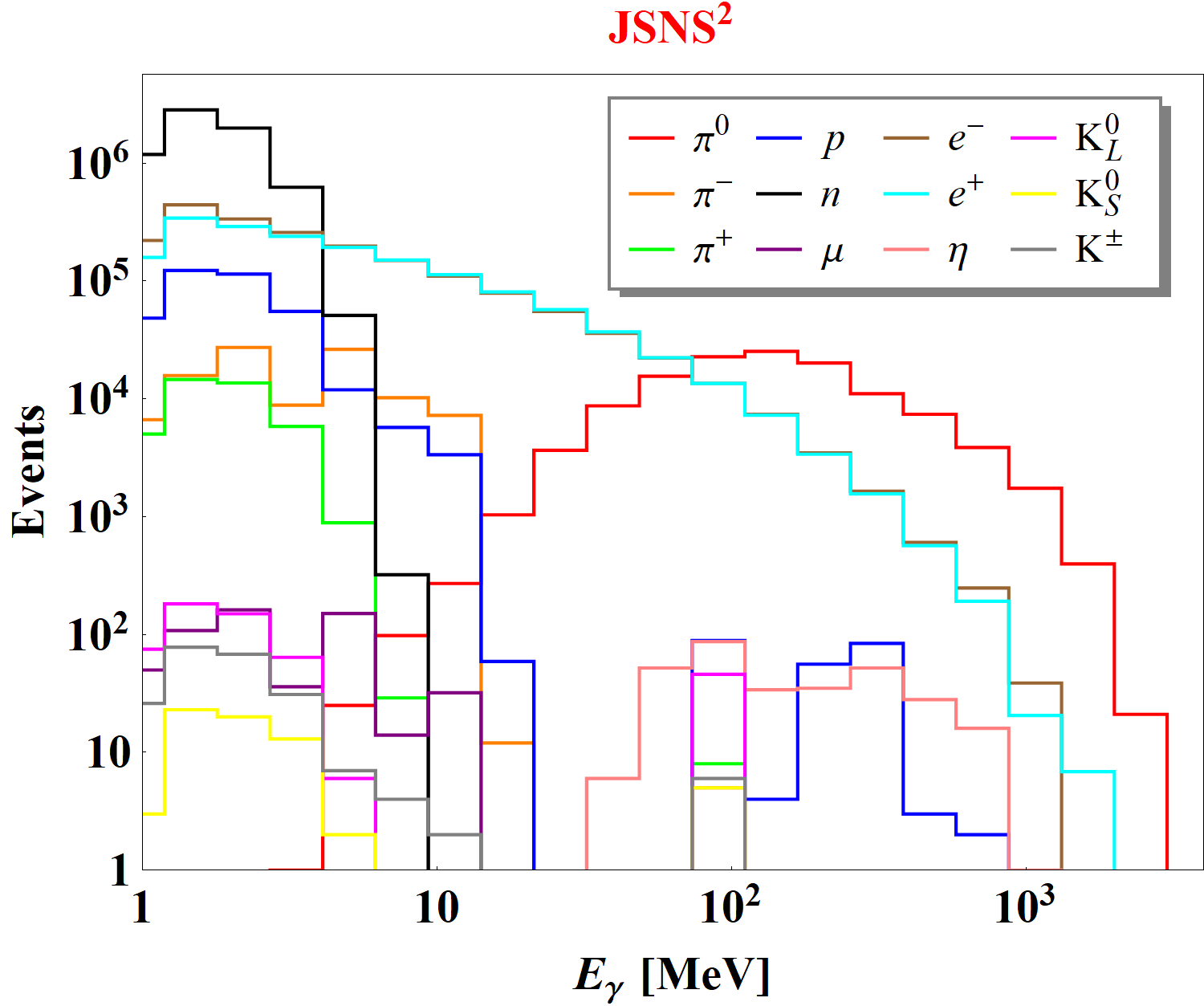}
    \includegraphics[width=7.4cm]{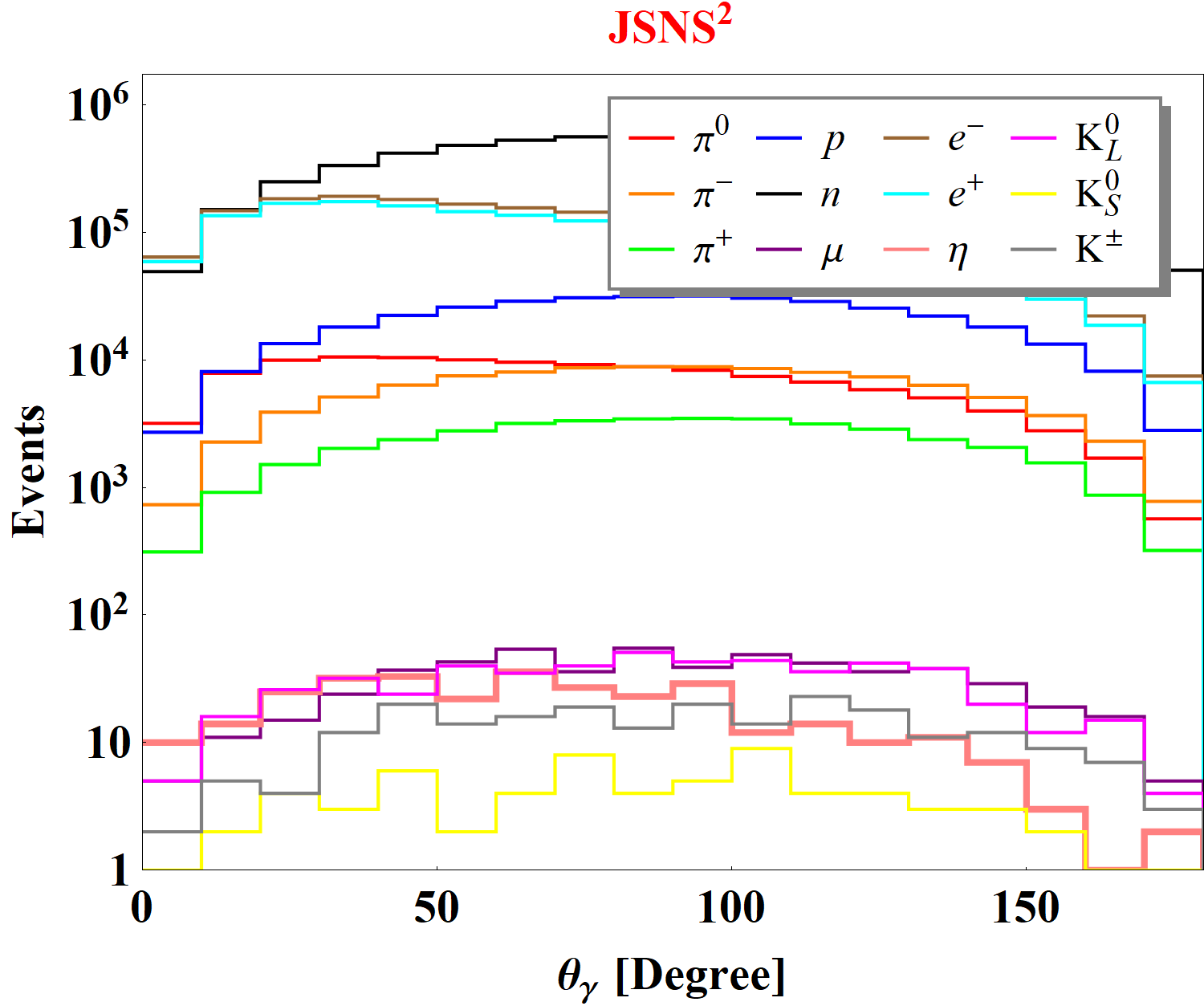} \\
    \vspace{0.3cm}    
    \includegraphics[width=7.4cm]{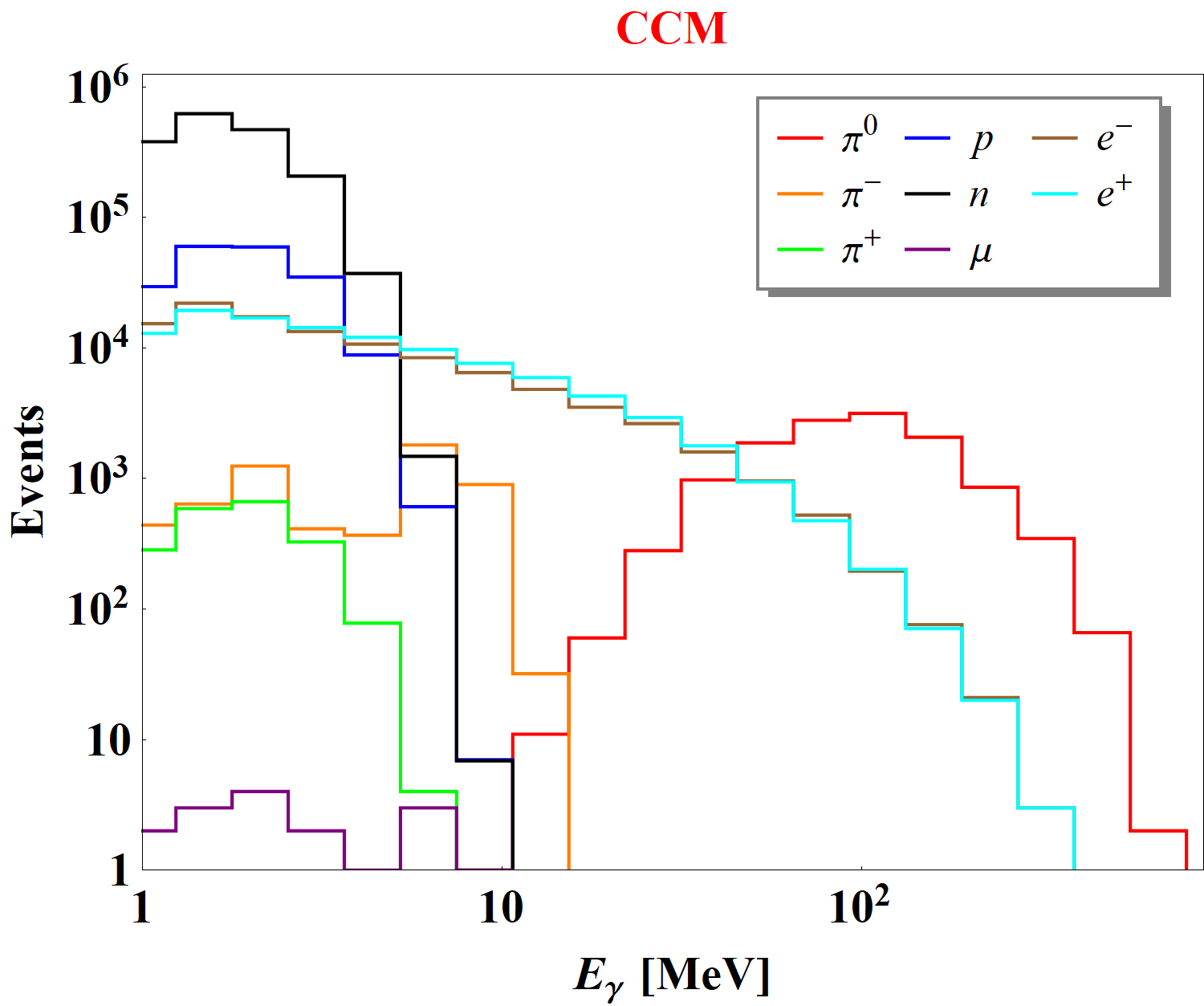}
    \includegraphics[width=7.4cm]{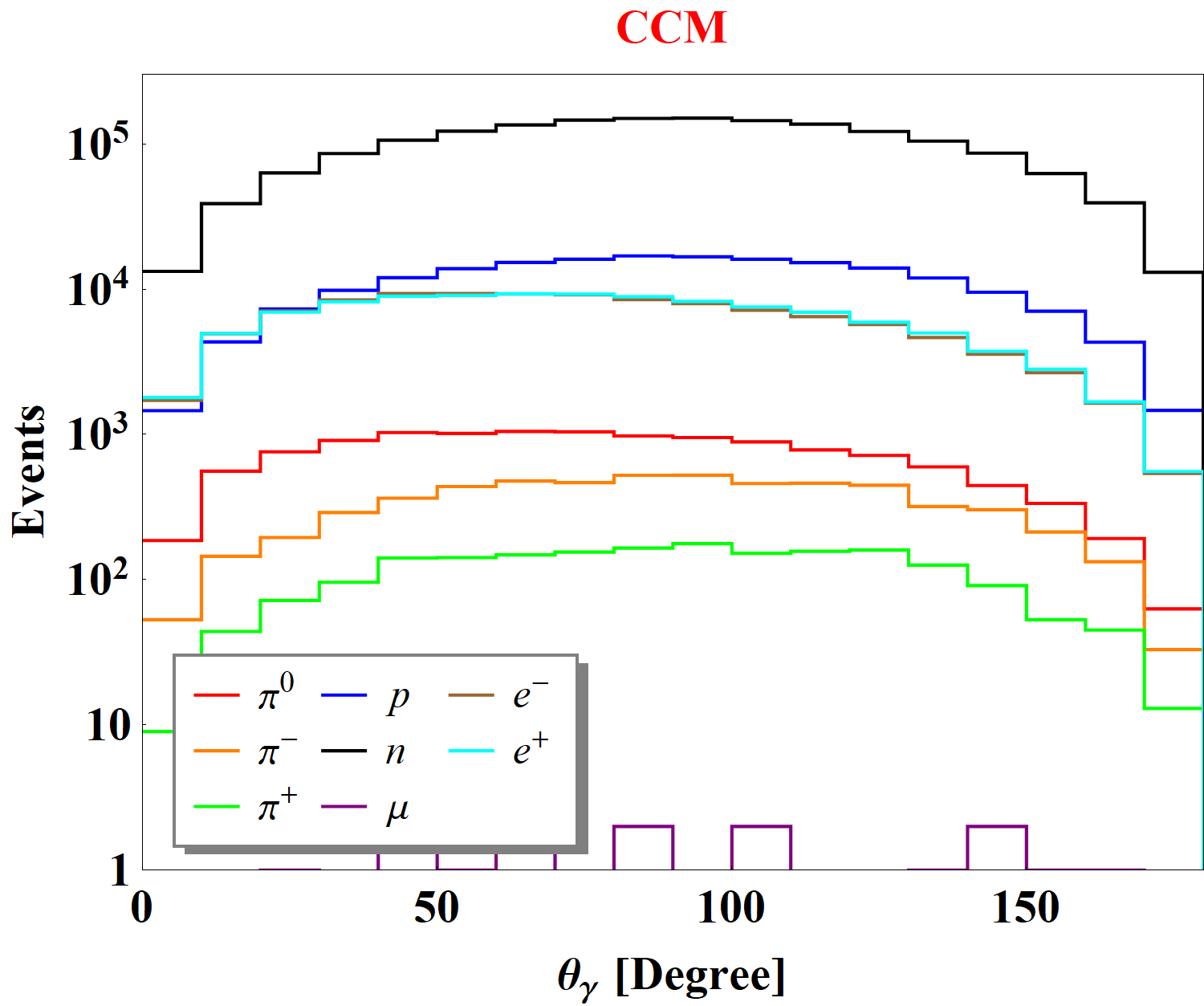}
    \caption{Photon energy (left) and angular (right) spectra in COHERENT (top), JSNS$^2$ (middle), and CCM (bottom) experiments according to various sources of photons.
    We use \texttt{GEANT}4 10.5 with the FTFP\_BERT library~\cite{Agostinelli:2002hh} for our simulation, generating $10^5$ protons incident on the targets. 
    The angle is measured from the incident proton beam line.} 
    \label{fig:photonspec}
\end{figure}

For operational convenience, our simulation study takes a pragmatic approach of converting an on-shell photon from the \texttt{GEANT} simulation to a dark gauge boson: the energy and the momentum direction of the dark gauge boson are the same as those of the photon, while the dark gauge boson flux is suppressed by $p_X^2$ with $p_X$ defined as a suppression factor for $\pi$ and $\eta$-related contributions.\footnote{The decay mode $\eta\to \gamma\gamma$ is about 40\%.
For the decay mode $\eta\to 3\pi^0$ ($\sim30\%$), the associated photon source is identified as $\pi^0$ not $\eta$ in our analysis.}
For cascade photons, the production of dark gauge bosons is facilitated by an off-shell electron, which brings about an additional phase space factor multiplying the suppression factor $p_X^2$, described in Appendix~\ref{sec:appA}.
The suppression factors according to the model assumption and the production mechanism are tabulated in Table~\ref{tab:supfactor}.
The suppression factors for model 2 of $T_{3R}$ case are shown in the parenthesis only if they differ compared to the model 1.
It is therefore important to estimate the energy and angular spectra of final-state photons as precisely as possible. 
We consider all available photon sources including not only the $\pi^0$ (and $\eta$ if available) decay and the $\pi^-$ absorption but cascade photons and neutron capture. 
Figure~\ref{fig:photonspec} exhibits photon energy and angular spectra in the left panels and the right panels, respectively.
The top panels, the middle panels, and the bottom panels are for the COHERENT, JSNS$^2$, and CCM experiments, correspondingly. 
As before, we generate $10^5$ protons on the target of each experiment to obtain these distributions, and the angle is measured with respect to the incident proton beam line. 

\begin{table}[t]
    \centering
    \resizebox{\columnwidth}{!}{%
    \begin{tabular}{c|c c c c c c c}
    \hline \hline 
    Production  & \multirow{2}{*}{$A'$} & \multirow{2}{*}{$B$} & \multirow{2}{*}{$L$} & \multirow{2}{*}{$B-L$} & \multirow{2}{*}{$L_e-L_\mu$} & \multirow{2}{*}{$L_e-L_\tau$} & \multirow{2}{*}{$T_{3R}$ [model 1,(2)]} \\
    mechanism  & & & & & & & \\
    \hline 
    $\pi^0$ decay  & $\epsilon^2$ & $\left(\frac{g_B/3}{e/3}\right)^2$ & $-$ & $\left(\frac{g_{B-L}/3}{e/3}\right)^2$ & $-$ & $-$ & $\left(\frac{2g_{T_{3R}}/3}{e/3}\right)^2$ \\
    $\pi^-$ absorption & $\epsilon^2$ & $\left(\frac{g_B/3}{2e/3}\right)^2$ & $-$ & $\left(\frac{g_{B-L}/3}{2e/3}\right)^2$ & $-$ & $-$ & $\left(\frac{2g_{T_{3R}}}{2e/3}\right)^2$\\
    $\eta$ decay  & $\epsilon^2$ & $\left(\frac{g_B/3}{e/3}\right)^2$ & $-$ & $\left(\frac{g_{B-L}/3}{e/3}\right)^2$ & $-$ & $-$ & $\left(\frac{2g_{T_{3R}}/3}{e/3}\right)^2$ \\
    $e^\pm$ cascade  & $\epsilon^2$ & $-$ & $\left(\frac{g_{L}}{e}\right)^2$ & $\left(\frac{g_{B-L}}{e}\right)^2$ & $\left(\frac{g_{L_e-L_\mu}}{e}\right)^2$ & $\left(\frac{g_{L_e-L_\tau}}{e}\right)^2$ & $\left(\frac{2g_{T_{3R}}}{e}\right)^2$, $(-)$\\
    \hline \hline
    \end{tabular}
    }
    \caption{Values of suppression factor $p_X^2$ according to the model assumption and the production mechanism. 
    $e$ is the ordinary electromagnetic gauge coupling constant.
    The ``$-$'' symbol is for the cases where the associated (tree-level) dark gauge charge $x_f^X$ is zero.  }
    \label{tab:supfactor}
\end{table}

For all three experiments, photons from most of the sources carry energies less than $\sim 5$~MeV.
As we will discuss in Section~\ref{sec:analysis}, low-energy dark matter does not deposit enough energy to overcome our energy cut or threshold in the nucleus recoil and the electron recoil. 
Indeed, the dark matter coming from the decay of dark gauge boson of $\lesssim 5$~MeV is the case. 
Therefore, photons from $\pi^0$ decays (and $\pi^-$ absorption that we will discuss shortly) make dominant contributions to our dark matter signal, followed by $e^\pm$-induced cascade photons. 
In the case of COHERENT and JSNS$^2$, we find that photons from the $\eta$ meson decay becomes the dominant source to produce a dark matter flux for $m_{\pi^0} <m_X < m_\eta$, while $e^\pm$-induced cascade photons make a subdominant contribution because of the suppression factors discussed above.  
Furthermore, we see that these photons are mildly more populated in the forward regime of angular spectra, as exhibited in the right panels of Figure~\ref{fig:photonspec}.
We therefore expect that detectors in the forward direction can receive a slightly more dark matter flux from the two sources.

Finally, let us discuss the simulation for the $\pi^-$ absorption process.
As argued earlier, the monochromatic Panofsky photons interact with nearby nucleons and electrons in a complex atom and there remain a number of soft photons in the final state (see also orange histograms in the left panels of Figure~\ref{fig:photonspec}). 
In order to estimate the dark matter flux from the $\pi^-$ absorption process, we develop our own simulation code to take care of the single-energy-valued dark gauge boson production, not relying on the \texttt{GEANT}4 package. 
The energy spectrum of these dark gauge bosons follows a Gaussian shape whose mean value is
\begin{equation}
    E_{X} = \frac{(m_p+m_{\pi^-})^2 -m_n^2 +m_X^2}{2(m_p+m_{\pi^-})}\,,
\end{equation}
and sigma value is 6\% as suggested in measurement data~\cite{Panofsky:1950he,MacDonald:1976ky}. 
Note that our \texttt{GEANT}4 simulation shows that most of the produced $\pi^-$ do not result in $\pi^0$ (see Table~\ref{tab:sim}), so we can assume that they are absorbed. 
We therefore make an approximation that the dark gauge boson flux is scaled by $p_X^2$ with respect to the full $\pi^-$ flux.

\section{Timing Spectrum of Dark Matter Events \label{sec:timing}}

The timing information of dark matter events is useful for discriminating them from potential backgrounds, in particular, delayed neutrino events. 
We begin with investigating expected timing spectra of simulated dark matter events, while providing the theoretical derivation of the differential timing spectrum in Appendix~\ref{sec:appB} for interested readers. 
Figure~\ref{fig:timespec} shows several representative unit-normalized timing spectra of the dark matter events in our benchmark experiments, COHERENT (top-left), CCM (top-right), and JSNS$^2$ (bottom). 
For JSNS$^2$, the timing spectra are plotted with respect to two consecutive pulses.
We consider three different pairs of the rest-frame lifetime of dark gauge boson $\tau_X$ and its mass $m_X$: $(m_X,\tau_X)=(75~{\rm MeV}, \leq 1~{\rm ns})$, $(75~{\rm MeV}, 1~\mu{\rm s})$, and $(138~{\rm MeV}, 1~\mu{\rm s})$ shown by blue, red, and orange histograms, respectively.
The first two points are relevant to relativistic dark gauge boson scenarios, while the last one invokes production of non-relativistic dark gauge boson via the $\pi^-$ absorption process. 
In terms of lifetime, the dark gauge boson in the first point decays rather promptly compared to the other two. 
The simulation sample for each scenario contains events contributed by production mechanisms P1 through P4 discussed in Section~\ref{sec:production}.
For comparison, we show the timing spectra of prompt neutrino and delayed neutrino events by the green and brown histograms, respectively, with neutrino scattering cross sections convolved.\footnote{In the case of JSNS$^2$, delayed neutrinos get an enhancement of charged-current interactions from $\nu_e$ scattering off electrons over their $\nu_\mu$ prompt counterparts, so their relative fractions are different from those in COHERENT and CCM.}
They are stacked and collectively unit-normalized.

\begin{figure}[t]
    \centering
    \includegraphics[width=7.5cm]{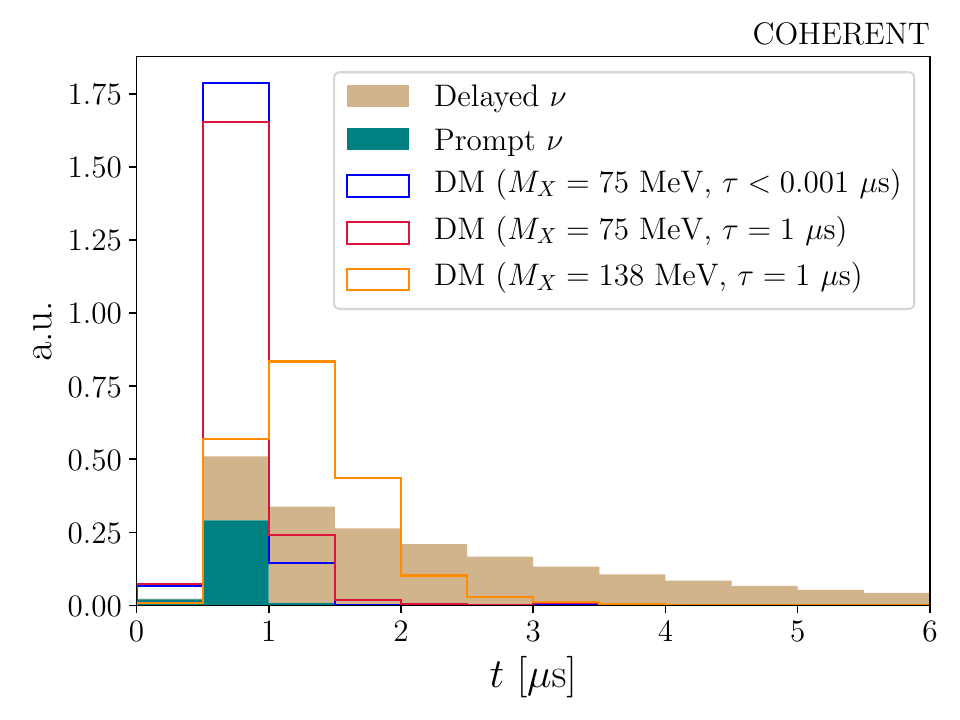}
    \includegraphics[width=7.5cm]{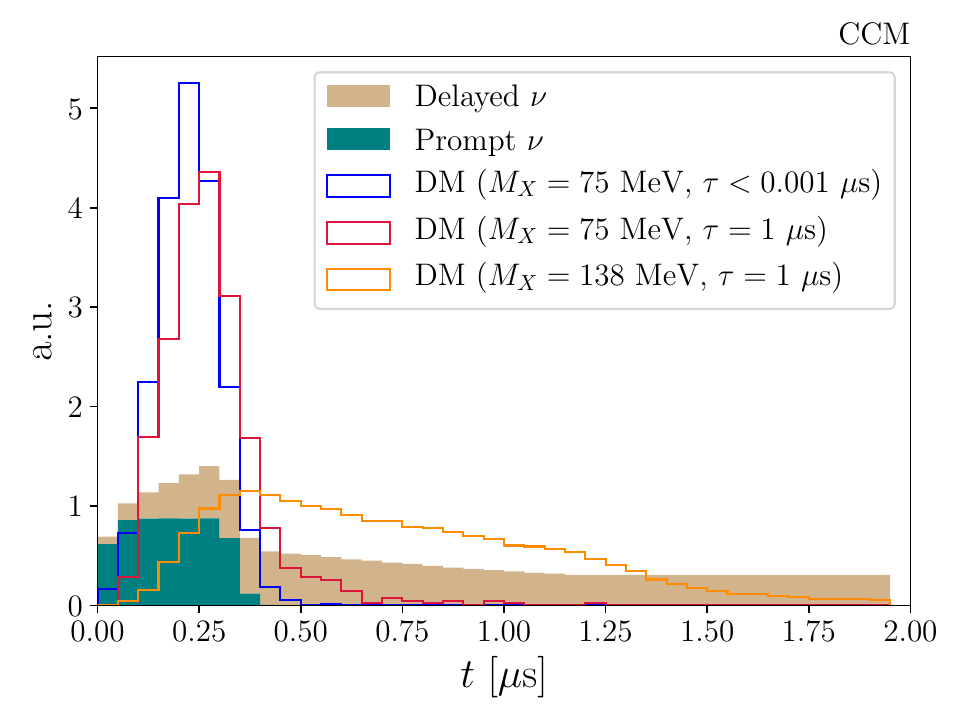} \\
    \includegraphics[width=7.5cm]{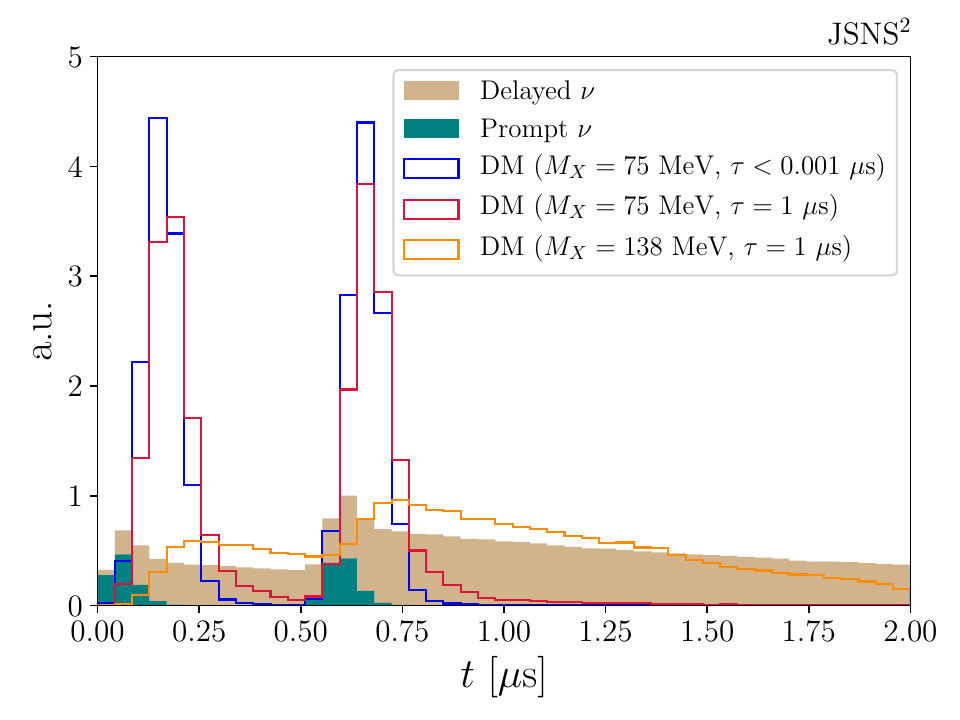}
    \caption{Expected unit-normalized timing distributions of simulated dark matter and neutrino scattering events contributed by production mechanisms P1 through P4 discussed in Section~\ref{sec:production}, at COHERENT (top-left), CCM (top-right), and JSNS$^2$ (bottom).
    The timing distributions for JSNS$^2$ are plotted for both the first and second pulses, taking the distribution from Ref.~\cite{Ajimura:2017fld}.
    We consider a long-lived ($\tau=1$ $\mu$s) and short-lived ($\tau<0.001$ $\mu$s) dark gauge boson at $75$ and $138$ MeV.
    For comparison, we show the timing spectra of prompt neutrino and delayed neutrino events by the green and brown histograms, respectively, with neutrino scattering cross sections convolved.
    They are stacked and collectively unit-normalized.
    }
    \label{fig:timespec}
\end{figure}

We observe two important features from this exercise. 
First, the dark matter flux reaching the detector gets maximized if the produced dark gauge boson decays rather promptly. 
If the lifetime of $X$ is too large, the decay point is too distant from the detector location so that the resulting dark matter flux is suppressed by distance square. 
Indeed, the dark matter flux remains (almost) maximized as long as dark gauge bosons flying toward the detector decay before reaching the detector. 
Therefore, if the laboratory-frame lifetime $\tau_X^{\rm lab}$ is smaller than $\sim 50-100$~ns, no significant loss of the dark matter flux is expected.
Otherwise, the dark matter flux is significantly dropped unless dark gauge bosons are produced almost at rest in which the timing spectrum can develop a sizable tail just like the delayed neutrinos. 
Assuming that the typical Lorentz boost factor of $X$ is 10 and BR$_{X\to 2\chi}\approx 1$, we find that $\kappa_D^X \gtrsim 10^{-6} - 10^{-7}$ for $m_X \in (10,~500)$~MeV.
Second, the simulation results show that most of the dark matter events are populated in certain timing windows. 
The upper limit in each window is closely related to the duration of a single beam pulse (roughly twice the beam full-width).
This will give a guidance for determining the timing cut for each benchmark experiment.
On the other hand, the lower limit reflects the required amount of time for a dark matter particle to arrive at the detector from the triggering moment of timing measurement, i.e., $t=0$.\footnote{Note that the choice of $t=0$ does not affect our analyses because one could simply shift the timing window.}
For COHERENT, due to processing and propagation delays of the POT signal, the timing of the POT signal effectively shows an arbitrary offset from signals in a detector which is experimentally determined by an {\it in-situ} measurement of prompt neutron signals.
Our COHERENT plot starts from $\sim 0.3~\mu$s which is based on the offset from Ref.~\cite{Akimov:2018vzs} even if the chosen bin size 0.5~$\mu$s is bigger than that. 
By contrast, all activities during $0-1~\mu$s will be recorded in JSNS$^2$, using the ``beam kicker'' timing~\cite{privateJSNS2}.
Thus, the lower limit in the JSNS$^2$ plot is set to be 0.
In a similar way, we set the lower limit in the CCM plot to be 0~\cite{privateCCM}.

\section{Data Analysis \label{sec:analysis}}

In this section, we discuss how to use the timing spectra in terms of new physics searches at neutrino experiments, in conjunction with recoil energy spectra. 
As we have discussed in Section~\ref{sec:detectiondm}, the dark matter particles that are produced manifest themselves as nucleus or electron recoil. 
Given this experimental signature, any SM neutrinos that reach the detector can appear dark matter signal-like. 
As briefly mentioned in Section~\ref{sec:experiments}, there are two types of neutrinos, ``prompt'' and ``delayed'' neutrinos. 

The former class of neutrinos are from the decay of charged pions, 
\begin{equation}
    \pi^\pm \to \mu^\pm + \nu_\mu\,. \label{eq:promptnu}
\end{equation}
Since the beam energies of our benchmark experiments are not large, the produced charged pions are not much energetic so that they quickly lose their available kinetic energy mainly by ionization and stop in the target material before decaying. 
Indeed, our \texttt{GEANT}4 simulation study suggests that decay-in-flight of $\pi^\pm$ be negligible, and this observation is supported by the dedicated Monte Carlo studies conducted by the JSNS$^2$ Collaboration~\cite{Ajimura:2017fld}.
Obviously, these neutrinos coming from the two-body decay process of $\pi^\pm$ nearly at rest are monochromatic. 
More importantly, $\mu^\pm$ carries away a dominant fraction of the $\pi^\pm$ rest-mass energy, leaving a small fraction to $\nu_\mu$, i.e., $E_{\nu_\mu}\approx 29.8$~MeV.
Therefore, the recoil energy deposited by such neutrinos is {\it bounded} by the upper kinematic limit. 
We shall shortly see that the energy of nucleus or electron recoil induced by these neutrinos is less than a definite value. 
For a given incoming neutrino energy $E_\nu$, we find that the maximum  kinetic energy of recoil nucleus and recoil electron, $E_{r,N}^{\max}$ and $E_{r,e}^{\max}$ are given by
\begin{eqnarray}
    E_{r,N}^{\max} &=& \frac{2E_{\nu}^2}{m_N+2E_{\nu}}\,, \\
    E_{r,e}^{\max} &=& \frac{2E_\nu^2 + 2E_\nu m_e +m_e^2}{m_e+2E_\nu}\,, \label{eq:ermaxelec}
\end{eqnarray}
where $m_N$ and $m_e$ are the mass of the target nucleus and the mass of the target electron, respectively, as defined earlier.
Note that for $E_\nu \gg m_e$, $E_{r,e}^{\max} \gg m_e$ is satisfied, so we quote the total energy in Eq.~\eqref{eq:ermaxelec} as the kinetic energy converges to the total energy in the relativistic regime.

On the other hand, the latter class of neutrinos are from the decay of muons that are produced through the process in Eq.~\eqref{eq:promptnu},
\begin{equation}
    \mu^\pm \to e^\pm + \nu_\mu +\nu_e\,. \label{eq:delayednu}
\end{equation}
The muons are not as energetic as the charged pions, so they quickly stop flying and then decay. 
However, muons are much longer-lived than charged pions by about two orders of magnitude (i.e., $\tau_\mu \approx 2.2~\mu$s vs. $\tau_{\pi^\pm} \approx 26$~ns). 
Therefore, the arrival times of the neutrinos in Eq.~\eqref{eq:delayednu} at the detector are much more {\it delayed} than those in Eq.~\eqref{eq:promptnu}.  
The energy of these neutrinos is not single-valued as they come from a three-body decay process, while it is still upper-bounded as muons decay nearly at rest. 
However, the neutrinos here can carry an energy up to $E_\nu\approx m_\mu/2 = 52.5$~MeV so that the resultant recoil energy spectrum can be more broadly distributed.

\begin{figure}
    \centering
    \includegraphics[width=7.5cm]{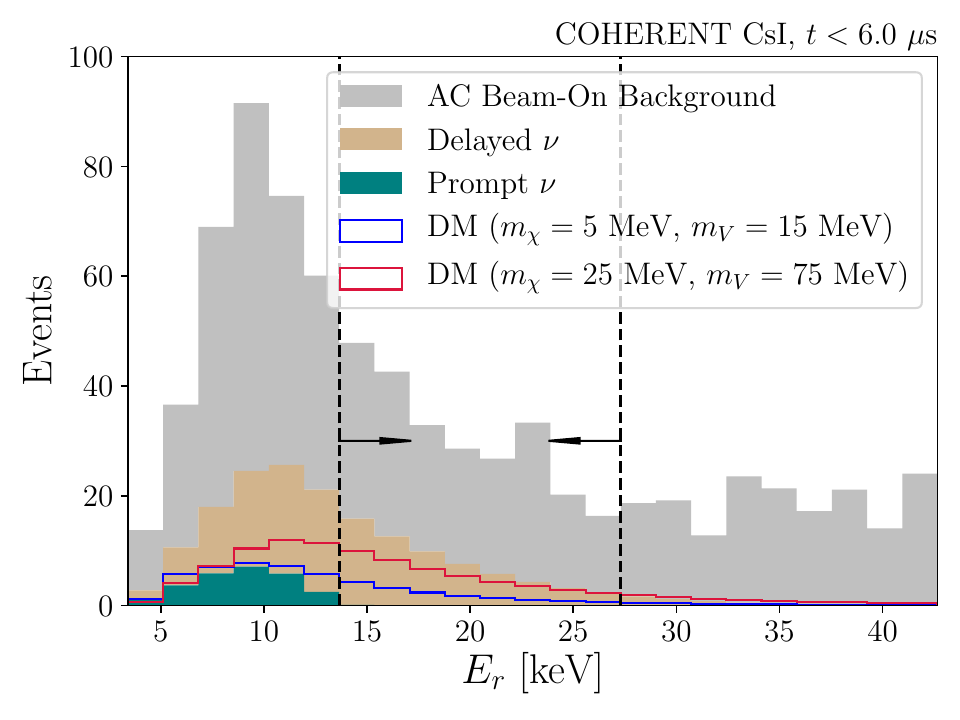}
    \includegraphics[width=7.5cm]{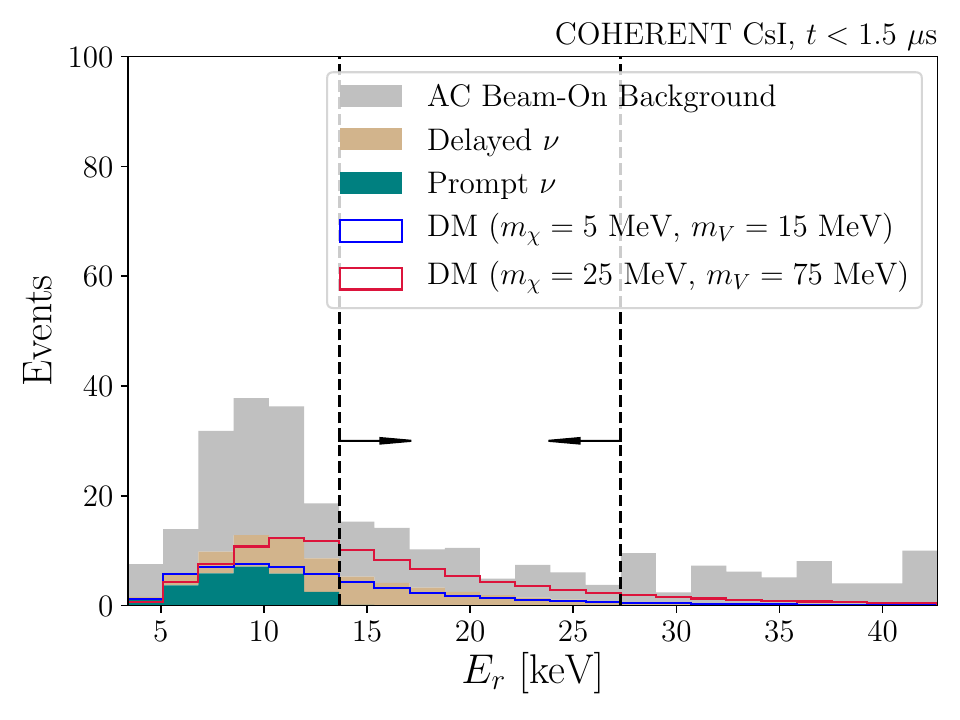}\\
    \includegraphics[width=7.5cm]{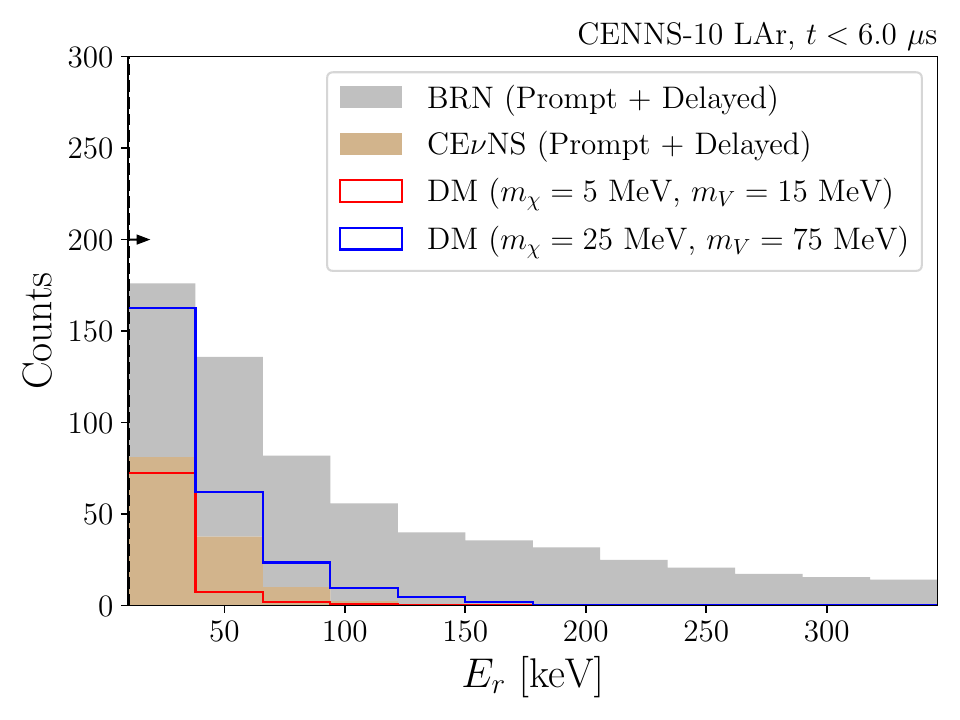}
    \includegraphics[width=7.5cm]{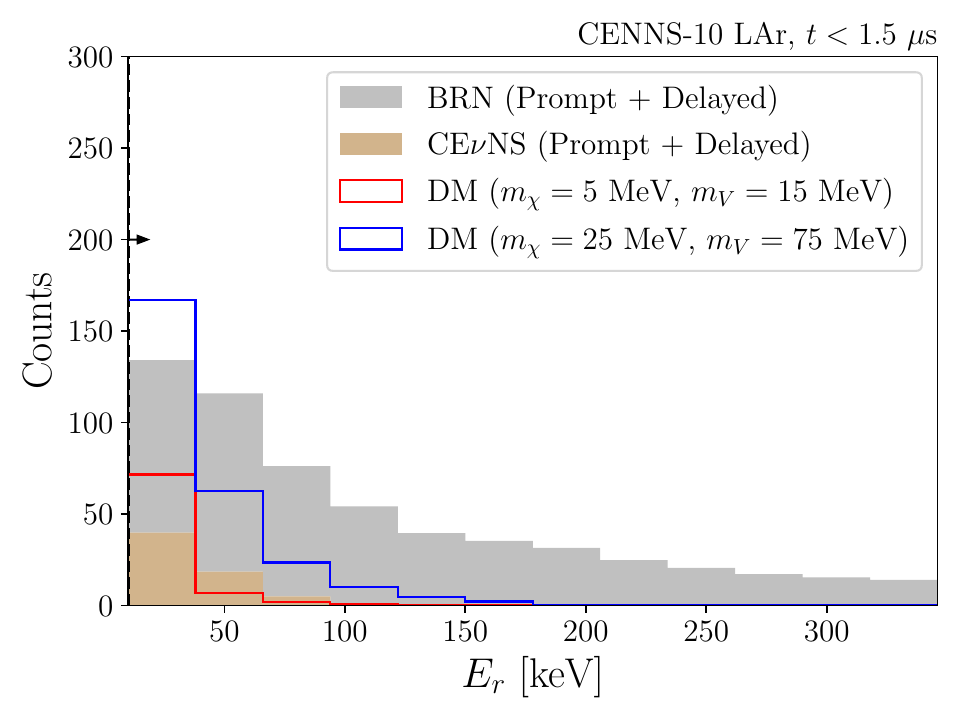}\\
    \caption{Recoil energy spectra produced from neutrino and dark matter interactions with nuclei in the CsI (top) and LAr (bottom) detectors. Spectra are shown before the timing cut (left) and after the timing cut (right).
    The vertical dashed lines indicate the energy cuts that are used to eliminate prompt $\nu$-induced events. Dark matter coherent scattering spectra are also shown for two choices of dark matter mass and mediator mass (up to an arbitrary choice of coupling).
    The AC background at the CsI detector includes BRN and SS, while the SS background at the CENNS-10 detector has been subtracted above.
    }
    \label{fig:spectra_coherent}
\end{figure}

\begin{figure}
    \centering
    \includegraphics[width=7.5cm]{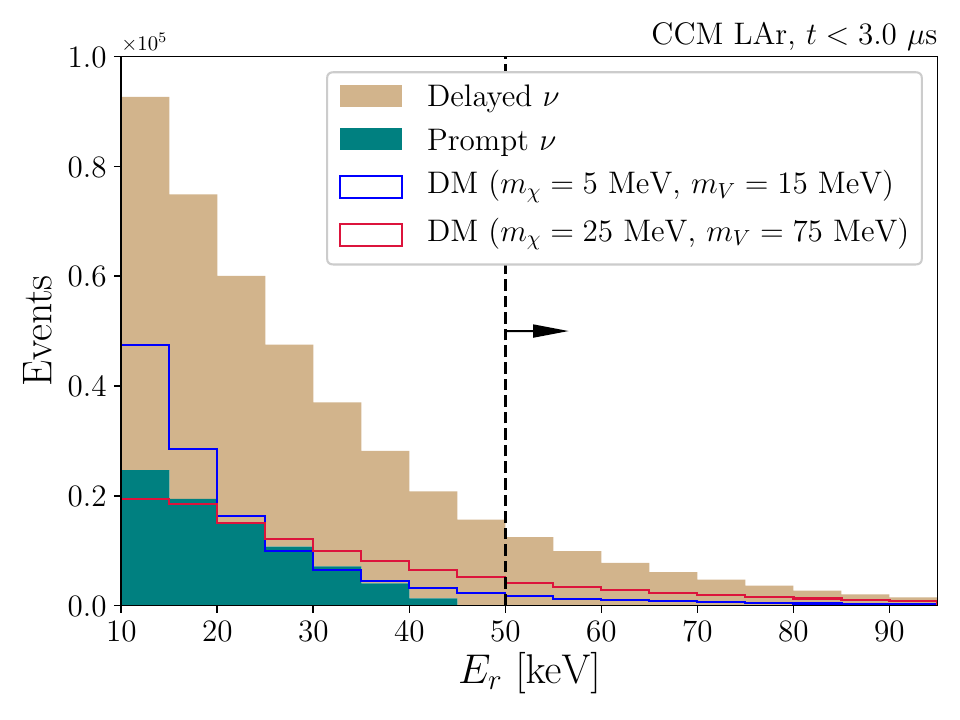}
    \includegraphics[width=7.5cm]{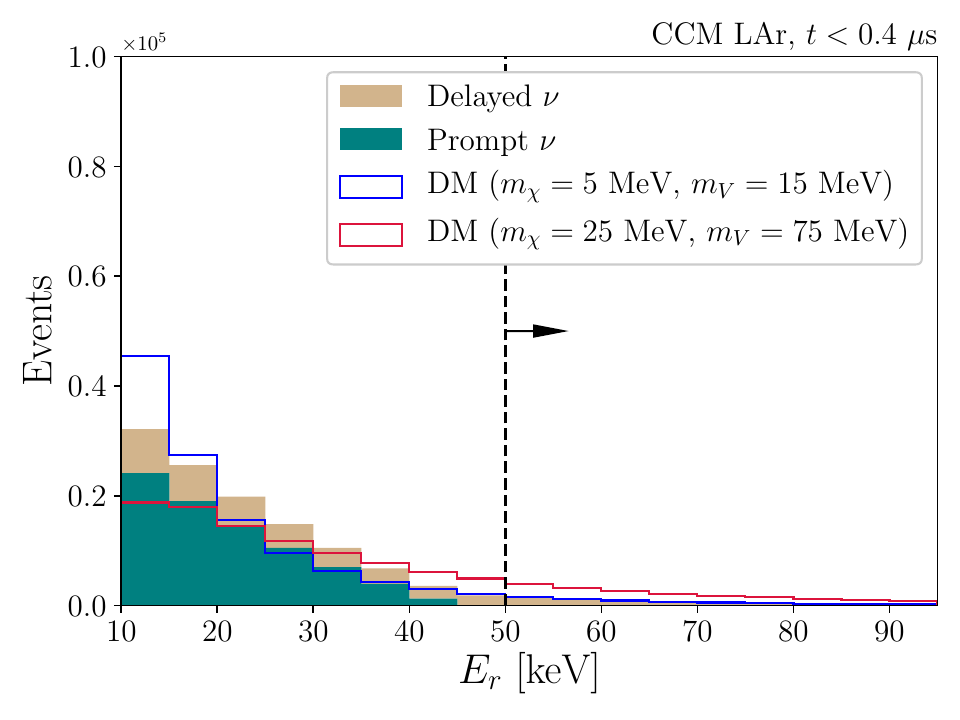} \\
     \includegraphics[width=7.5cm]{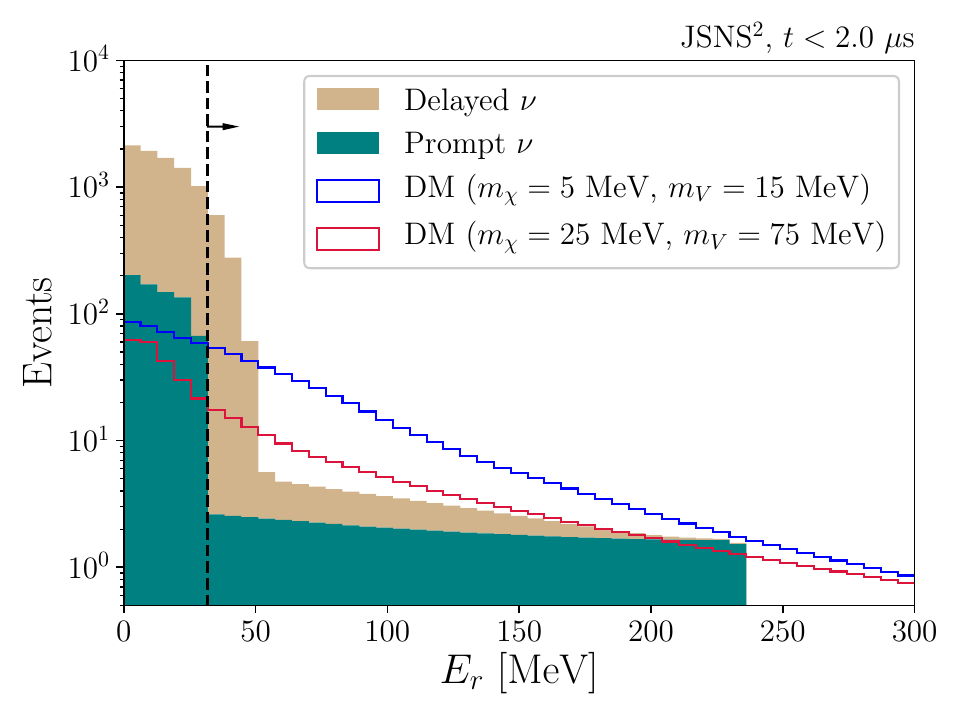}
    \includegraphics[width=7.5cm]{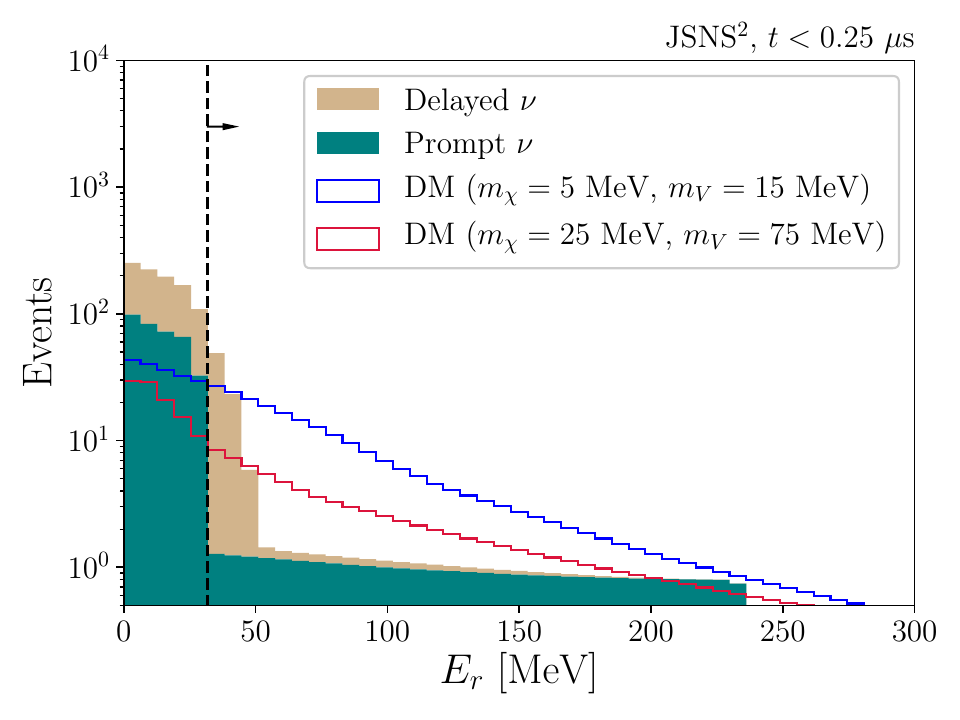}
    \caption{Recoil energy spectra produced from neutrino and dark matter interactions with nuclei in the LAr detector at CCM (top), and those with electrons in the Gd-LS detector at JSNS$^2$ (bottom).
    For JSNS$^2$, included are prompt neutrinos from kaon decays at rest whose energy spectrum falls off sharply at $E_r\approx 235$~MeV. No detector efficiencies or energy resolution smearing effects were assumed.
    Spectra are shown before the timing cut (left) and after the timing cut (right).
    The vertical dashed lines indicate the energy cuts that are used to eliminate prompt $\nu$-induced events.
    Dark matter coherent scattering spectra are also shown for two choices of dark matter mass and mediator mass (up to an arbitrary choice of coupling).
    }
    \label{fig:spectra}
\end{figure}

Given these features of the prompt and delayed neutrino-induced background events, we propose to apply a combination of an energy cut and a timing cut in order to suppress the SM neutrino backgrounds but retain as many dark matter signal events as possible.\footnote{The idea of using the timing and energy cuts was adopted in the context of beam-focusing experiments, e.g., MiniBooNE~\cite{MiniBooNE:2012jpi}, where massive dark matter would be slightly less relativistic than neutrinos and the majority of dark matter events would populate in delayed timing bins.}
The main ideas behind the proposed selection scheme can be summarized as follows. 

\medskip

\noindent {\bf Energy cut}:
The background events produced from the prompt neutrino populate below a certain value in the recoil energy spectrum, whereas signal events can deposit larger energy through the recoil as incoming dark matter particles are typically more energetic than the prompt neutrinos. 
An appropriately chosen energy cut can, therefore, suppress prompt $\nu$-induced events substantially while a large fraction of signal events still survive. 
The left panels in Figure~\ref{fig:spectra_coherent} exhibit energy spectra of background events and signal events in the CsI (top) and the LAr (bottom) detectors of COHERENT.
The backgrounds include not only prompt-neutrino events (green) and delayed-neutrino events (brown) but other backgrounds such as beam-related neutron (BRN) events (gray) as reported in Refs.~\cite{Akimov:2018ghi,Akimov:2020pdx}. 
The anti-coincidence (AC) beam-on data at the CsI detector of COHERENT includes the steady-state (SS) background as well as BRN. 
The dark matter spectra are shown for two choices of dark matter mass and mediator mass (blue and red) modulo an arbitrary choice of coupling. 
These energy spectra clearly demonstrate the expectation that a sizable fraction of dark matter events populate beyond the endpoint of the prompt-neutrino energy spectrum.
Similar behaviors in the energy spectra are expected for CCM and JSNS$^2$ and they are shown in the left panels of Figure~\ref{fig:spectra}. 
For JSNS$^2$, included are prompt neutrinos from kaon decays at rest whose energy spectrum falls off sharply at $E_r\approx 235$~MeV.
As no BRN and SS backgrounds are available for them yet, we compare dark matter events only with neutrino-induced background events.  
Since the JSNS$^2$ Gd-LS detector has a large energy threshold (2.6~MeV), the nuclear scattering channel may not be available so that we display the energy spectra of recoiling electrons for the Gd-LS case. 

We clearly see that the application of an energy cut (denoted by the vertical dashed lines) can eliminate prompt $\nu$-initiated events very efficiently.
Inspired by this observation and based on our cut optimization, we choose the following selection criteria:
    \begin{equation}
        E_r > \left\{
        \begin{array}{l l}
            14~{\rm keV} 
            & \hbox{ for CsI of COHERENT (nucleus scattering)} \\
            10~{\rm keV} & \hbox{ for LAr of COHERENT (nucleus scattering)} \\
            50~{\rm keV}  &  \hbox{ for LAr of CCM (nucleus scattering)}\\ 
            30~{\rm MeV} & \hbox{ for Gd-LS of JSNS$^2$ (electron scattering),} 
        \end{array}\right.\label{eq:energycut}
    \end{equation}
which are also indicated by the rightward arrows in Figures~\ref{fig:spectra_coherent} and \ref{fig:spectra}. 
For the CsI detector of COHERENT, we further apply an upper energy cut beyond which background uncertainties are high~\cite{Akimov:2018ghi}.
\begin{equation}
    E_r < 26~{\rm keV}~~~\hbox{for CsI of COHERENT (nucleus scattering)}.\label{eq:energyupcut}
\end{equation}
This selection is indicated by the leftward arrow in the top panels of Figure~\ref{fig:spectra_coherent}. 
Our choices for the energy cut are summarized in Table~\ref{tab:cutsummary}.

Two comments should be made in order.
First, as shown by the blue and the red histograms, the above cut choices are not necessarily optimized to all possible mass points. 
Our choices are based on the optimization with dark gauge boson being $\sim 100$~MeV and much lighter dark matter merely for illustration, and we do not perform an optimization procedure mass point-by-point. 
Second, we see that a large fraction of delayed $\nu$-induced events (brown histograms) survive after the energy cut, as discussed in the previous section. 
This motivates us to introduce a timing cut to reject them further.

\begin{table}[t]
    \centering
    \resizebox{\columnwidth}{!}{
    \begin{tabular}{c|c c c | c}
    \hline \hline
         &  \multirow{2}{*}{Channel} & \multirow{2}{*}{$E_r$ cut} & \multirow{2}{*}{$t$ cut} & Background eff.\\
           &   &  &  & (Remaining events)\\
    \hline
    \multirow{2}{*}{COHERENT-CsI} & \multirow{2}{*}{Nucleus}  & \multirow{2}{*}{$E_r \in (14,26)$~keV} & \multirow{2}{*}{$t<1.5~\mu$s} & AC: 13.4\% (138) \\
     & & & & CE$\nu$NS: 37.9\% (69)\\
    \multirow{2}{*}{COHERENT-LAr} & \multirow{2}{*}{Nucleus}  & \multirow{2}{*}{$E_r>10$~keV} & \multirow{2}{*}{$t<1.5~\mu$s} & AC: 15\% (550) \\
     & & & & CE$\nu$NS: 21\% (27) \\
    \multirow{2}{*}{CCM} & \multirow{2}{*}{Nucleus} & \multirow{2}{*}{$E_r>50$~keV} & $t<0.1~\mu$s (Tight WP) & CE$\nu$NS: 0.03\% (189) \\
     & & & $t<0.4~\mu$s (Loose WP) & CE$\nu$NS: 0.88\% (5,970) \\
    JSNS$^2$ & Electron  & $E_r>30$~MeV & $t<0.25~\mu$s & CE$\nu$NS: 1.17\% (107) \\
    \hline \hline
    \end{tabular}
    }
    \caption{A summary of the recoil energy and timing cuts that we use for our data analysis. 
    The last column shows the background efficiencies and the remaining events after the cuts in the third and fourth columns. 
    The remaining events are based on 3 year exposures (JSNS$^2$ and CCM) and for 6576 kg$\cdot$days (COHERENT-LAr) and 4466 kg$\cdot$days (COHERENT-CsI).}
    \label{tab:cutsummary}
\end{table}

\medskip

\noindent {\bf Timing cut}:
The discussion in Section~\ref{sec:timing} suggests that a large portion of relativistic (non-relativistic) $X$-induced dark matter events irrespective of $\tau_{X}$ (with $\tau_{X} \lesssim 0.1~\mu$s) should be retained, as far as we keep the events in prompt timing bins. 
Again based on our cut optimization and private communications with experimentalists~\cite{private, privateCCM, privateJSNS2}, we have chosen a set of timing cuts:
    \begin{equation}
        t<\left\{
        \begin{array}{l l}
            1.5~\mu{\rm s} & \hbox{ for COHERENT} \\
            0.1~\mu{\rm s} & \hbox{ for CCM (Tight WP)} \\ 
            0.4~\mu{\rm s} & \hbox{ for CCM (Loose WP)} \\ 
            0.25~\mu{\rm s} & \hbox{ for JSNS$^2$,} 
        \end{array}\right. \label{eq:timecut}
    \end{equation}
and these are tabulated again in Table~\ref{tab:cutsummary}.
For CCM, we consider two working points (WP): a ``tight" cut at 0.1 $\mu$s, based on the experimental recommendation, and a ``loose" cut at 0.4 $\mu$s, based on the timing spectrum of the dark matter signal (see the top-right panel of Figure~\ref{fig:spectra}).
In particular, we will show the full power of the timing cut with this loose cut in Section~\ref{sec:interpretations}, and thus show a bigger potential of CCM in terms of the dark matter search. 
Most of delayed $\nu$-induced events reach the detector much later, as argued earlier. 
The right panels of Figures~\ref{fig:spectra_coherent} and \ref{fig:spectra} show the recoil energy spectra of signal and neutrino background events after applying the timing cuts in \eqref{eq:timecut}, and we see that most of delayed $\nu$-induced events (brown) can be rejected as compared to the corresponding left panels, allowing the signal events to stand out. 
We recall that the optimal choices for the timing cut depend on the duration of a single beam pulse.
As described in Section~\ref{sec:exp}, the JSNS$^2$ beam injects two consecutive $0.1~\mu$s wide pulses separated by an interval of 0.44~$\mu$s in each beam period.
We keep the duration of the first pulse as our baseline timing cut for JSNS$^2$, as the time window of the second pulse may be contaminated by delayed neutrino events generated by the first pulse. 

Finally, we report the background efficiencies after the cuts that we have proposed so far for each of the benchmark experiments in the last column of Table~\ref{tab:cutsummary}. 
The numbers of the remaining background events are estimated in the basis of three year exposures (JSNS$^2$ and CCM) and for 6576 kg$\cdot$days (COHERENT-LAr) and 4466 kg$\cdot$days (COHERENT-CsI).

\section{Interpretations \label{sec:interpretations}}

We are now in the position to discuss how to interpret the experimental results from our proposed analysis technique, in terms of new physics searches. 
In the first two subsections, we apply our event selection scheme defined in~\eqref{eq:energycut} and~\eqref{eq:timecut} for the experimental data collected at the CsI and LAr detectors of the COHERENT experiment~\cite{Akimov:2018ghi,Akimov:2020pdx}, and demonstrate a moderate excess. 
We then attempt to explain the moderate excess with a dark matter interpretation, assuming that it is real. 
By contrast, the remaining subsections are devoted to ways of constraining various dark matter models described in Section~\ref{sec:models}. 
For COHERENT, we assume that the excess could be explained by an unidentified background or a systematic uncertainty on the observed steady-state background, i.e., the number of observed events are consistent with the number of expected background events. 
We will present expected sensitivity reaches at the other benchmark detectors listed in Table~\ref{tab:expspec} under the assumption of null signal observations as well. 

As discussed in Section~\ref{sec:models}, the dark sector gauge boson responsible for production of dark matter can be different from the dark sector gauge boson that is exchanged in detection of dark matter.
We therefore consider two scenarios throughout this section. 
\begin{itemize}
    \item {\bf Single-mediator scenario}:
    In this case, $X^\mu=V^\mu$, so we have $\kappa_f^X=\kappa_f^V$, $\kappa_D^X=\kappa_D^V$, and $m_X=m_V$.
    Since the dark matter event rate at a detector is proportional to the dark matter flux times the detection cross section, an experiment obeying our search scheme is sensitive to an effective coupling $\kappa_{\rm eff}$, a combination of the coupling constants, given by 
    \begin{equation}
    \kappa_{\rm eff} \equiv \kappa_f^V \kappa_D^V \kappa_f^X \sqrt{{\rm BR}_{X\to \chi\bar{\chi}}(\kappa_D^X)} \rightarrow (\kappa_f^X)^2\kappa_D^X\sqrt{{\rm BR}_{X\to \chi\bar{\chi}}(\kappa_D^X)}\,, \label{eq:effcouple}
    \end{equation}
    where the $\kappa_D^X$ dependence is encoded in the branching ratio of $X$ to a dark matter pair. 
    In this scenario, relatively short-lived $X$, hence relatively large $\kappa_D^X$, are favored.
    Otherwise, the detection cross section would be too small to develop an enough sensitivity. 

    \item {\bf Double-mediator scenario}:
    By construction, the coupling constants and the mediator mass parameters can be different.
    Unlike the single-mediator scenario, relatively small $\kappa_D^X$, hence relatively long-lived $X$ are allowed because a sizable detection cross section is possible with a large $\kappa_D^V$. 
\end{itemize}

\subsection{Excess and dark matter interpretations - CsI Data \label{sec:explain}}
We revisit the analysis performed in our companion paper~\cite{Dutta:2019nbn}. 
We first review how the CsI data was analyzed based on the search strategy discussed in the previous section, and then interpret the result with a more complete set of dark matter sources. 
The data, as described in Ref.~\cite{Akimov:2018vzs}, includes observed signal ($N_{\rm obs, sig}$) and background ($N_{\rm obs, bg}$) counts.
We take the AC beam-on data from the 4466 kg$\cdot$day CsI run data at COHERENT as the observed background events $N_\text{obs,bg}$.
As mentioned earlier, this AC data represents SS and BRN backgrounds.
The relevant cuts in \eqref{eq:energycut}, \eqref{eq:energyupcut}, and \eqref{eq:timecut} were applied to the published COHERENT CsI data~\cite{Akimov:2018vzs} to reduce both prompt and delayed neutrino events.
Taking the experimental efficiencies given in Ref.~\cite{Akimov:2018vzs} into account and setting the baseline size of the neutron distribution $R_n$ of CsI to be 4.7~fm, we found that 97 events (denoted by $N_{\rm obs}=N_{\rm obs,sig}+N_{\rm obs,bg}$) pass the cuts.
Among them, 49 were classified as the SS background (denoted by $N_{\rm SS}$), 19 were identified as the neutrino-induced (denoted by $N_\nu$) background originating from the delayed neutrino, and 3 were classified as the BRN background (denoted by $N_{\rm BRN}$), resulting in 26 events left.
We take the definition of the significance in Ref.~\cite{Scholberg:2018vwg}:
\begin{equation}
    {\rm Significance} =\frac{N_{\rm obs}-N_{\rm SS}-N_\nu-N_{\rm BRN}}{\sqrt{2N_{\rm SS}+N_\nu+N_{\rm BRN}}}\,,
\end{equation}
from which these extra events correspond to a $\sim2.4\sigma$ deviation. 
For the analysis with $R_n=5.5$~fm~\cite{Cadeddu:2017etk}, the significance becomes $\sim 3.0\sigma$~\cite{Dutta:2019nbn}.

Given this ``excess'', we attempt to explain it by fitting the data to the dark matter model discussed in Section~\ref{sec:models}.
We assume that both the observed signal-like counts $N_{\rm obs,sig}$ and the observed background counts $N_{\rm obs,bg}$ follow Poisson models, where the Poissonian expectations are given by the ``true'' background\footnote{The background realized in the asymptotic limit.}
and signal counts, $N_{\rm bg}$ and $N_{\rm sig}$.
We parameterize the signal model counts as follows:
\begin{equation}
    N_{\rm sig}(t,E_r;\vec{\theta}) = (1+\alpha)\left\{N_{\rm \chi}(t,E_r;\vec{\theta}) + N_{\nu}(t,E_r)\right\}\,,
\end{equation}
where $\vec{\theta}$ is the vector of dark matter model parameters, and we additionally include the nuisance parameter $\alpha$ which controls the systematic uncertainties from the flux, nuclear form factor, quenching factor, and signal acceptance.
Notice here that we have defined our signal model as the inclusive sum of dark matter and neutrino events; this is purely a choice of formalism.
The timing and energy cuts that we introduced in Section~\ref{sec:analysis} are motivated to remove neutrino events that contaminate the signal model in this way.

With the definitions and assumptions above, we define a binned likelihood $\mathcal{L}(\mathcal{D}_{\rm CsI}\mid \vec\theta ;H_0)$ given the CsI data $\mathcal{D}_{\rm CsI}$, dark matter model parameters $\vec\theta$, and null hypothesis $H_0$ for which $\vec\theta = (0,\dots,0)$.
We can maximize this likelihood to determine the best-fit on $N_{\rm sig}$ (and therefore $\vec{\theta}$), by marginalizing over $\alpha$ and $N_{\rm bg}$ along all timing and energy bins $(t, E_r)$ as follows:
\begin{equation}
    \label{eq:L}
  \mathcal{L}(\mathcal{D}_{\rm CsI} \mid \vec\theta ; H_0)\propto \prod_{(t,E_r)}\int dN_{\rm bg} \int d\alpha P(N_\text{obs,sig},N_\text{sig})P(N_\text{obs,bg},N_\text{bg}) G(\alpha,\sigma_\alpha^2)\ ,
\end{equation}
where $P(k,\lambda)$ is the Poisson likelihood for observing $k$ events given a mean expectation $\lambda$, and $G(\alpha,\sigma^2_\alpha)$ is a Gaussian distribution over $\alpha$ with a mean of zero and variance of $\sigma_\alpha^2$.
For the COHERENT CsI data, the beam-on anti-coincidence data plays the role of $N_{\rm obs,bg}$, the beam-on coincidence data plays the role of $N_{\rm obs,sig}$, and $\sigma_\alpha$ is taken to be $0.28$~\cite{Akimov:2017ade}.

We consider the dark matter model in the presence of a single mediator (i.e., $X=V$), which is taken to be a dark photon ($A'$ in Table~\ref{tab:modelsummary}) for illustration, in performing a fit to the CsI data. 
This analysis is similar to the one performed in Ref.~\cite{Dutta:2019nbn}, with the important addition of a more complete set of sources for dark gauge boson production (P1 through P4 defined in Section~\ref{sec:production}).
We then set $m_V/m_\chi$ and $\alpha_D \equiv (\kappa_D^V)^2 / (4\pi)$ to be 3 and 0.5, respectively. 
With these fixed, the remaining free parameters are the mass $m_V$ and the kinetic mixing strength $\epsilon$.
For the sake of clarity in comparing our results to other analyses, we may freely reparameterize the coupling using the conventionally defined variable $Y\equiv \epsilon^2 \alpha_D (\frac{m_\chi}{m_V})^4$.

We use the Bayesian inference package \texttt{MultiNest}~\cite{Feroz:2008xx} which samples the parameter space $(m_V, Y)$ to evaluate the posterior probability distribution $p(m_V,Y \mid \mathcal{D}_\text{CsI})$ via Bayes' theorem:
\begin{equation}
    p(m_V, Y \mid \mathcal{D}_\text{CsI}) = \dfrac{\mathcal{L}(\mathcal{D}_\text{CsI} \mid m_V,Y ; H_0) \cdot \{\pi(m_V,Y)\}}{\mathcal{Z}},
\end{equation}
where $\mathcal{L}$ is the likelihood in Eq.~\eqref{eq:L}, now written in terms of model parameters $m_V$ and $Y$, $\pi(m_V,Y)$ is the uniform prior density over the appropriate ranges of $Y$ and $m_V$, and $\mathcal{Z}$ is the Bayesian evidence.\footnote{In practice, we pass in the logarithm of the likelihood $\mathcal{L}$ to \texttt{MultiNest}.
Additionally, the variables $Y$ and $m_V$ are drawn in log space from the inverse cumulative distribution fields of $\pi(m_V, Y)$ to hasten the conversion of the evidence calculation.}

\begin{figure}[t]
    \centering
    \includegraphics[width=7.5cm]{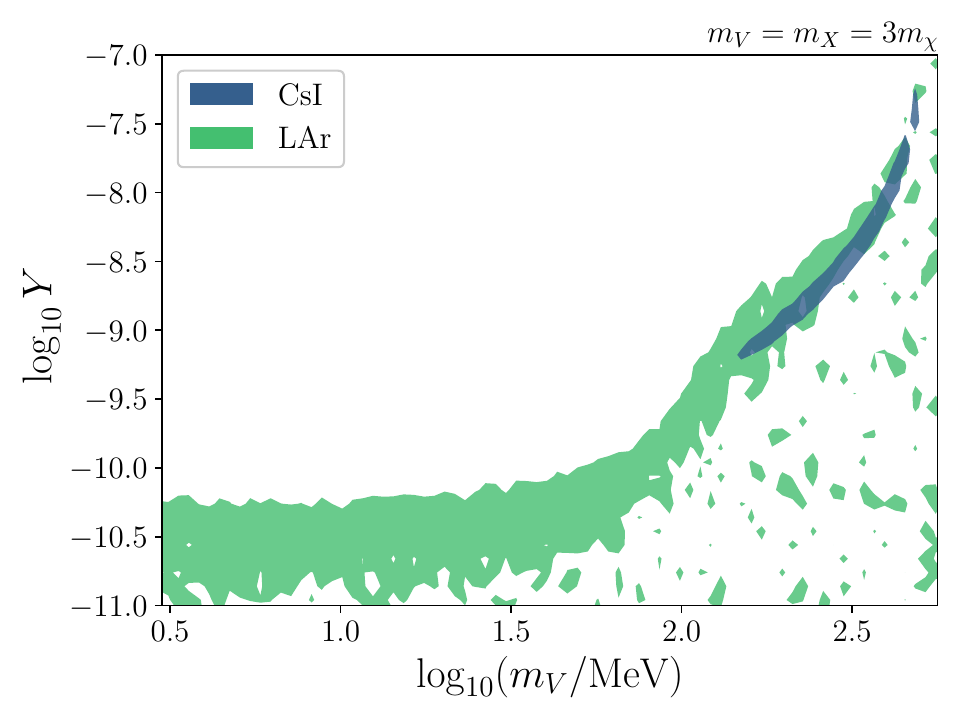} \\
    \includegraphics[width=7.5cm]{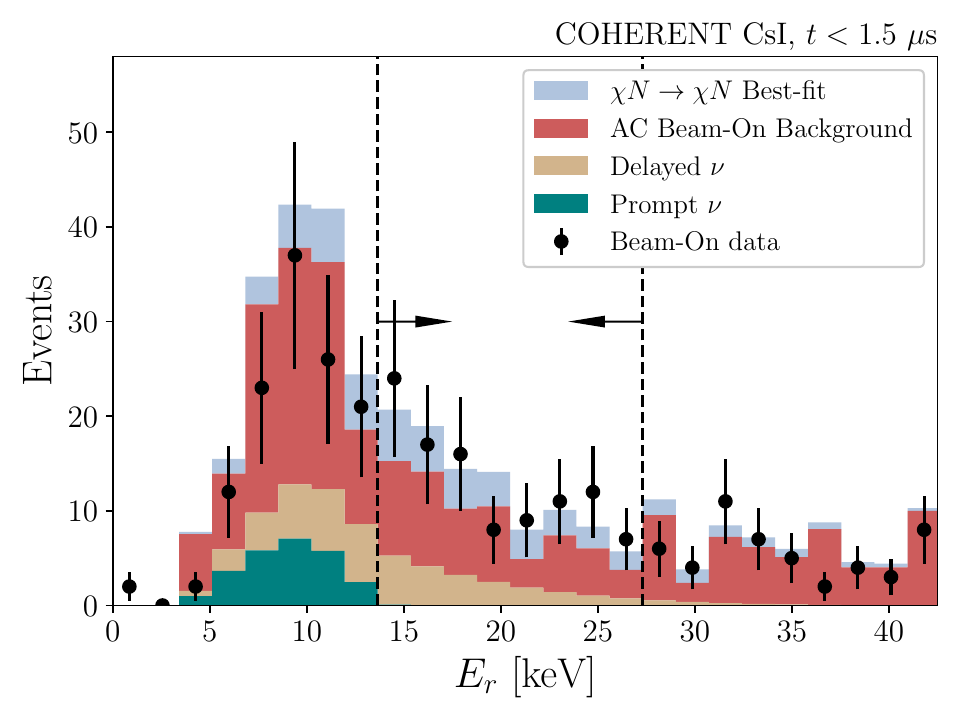}
    \includegraphics[width=7.5cm]{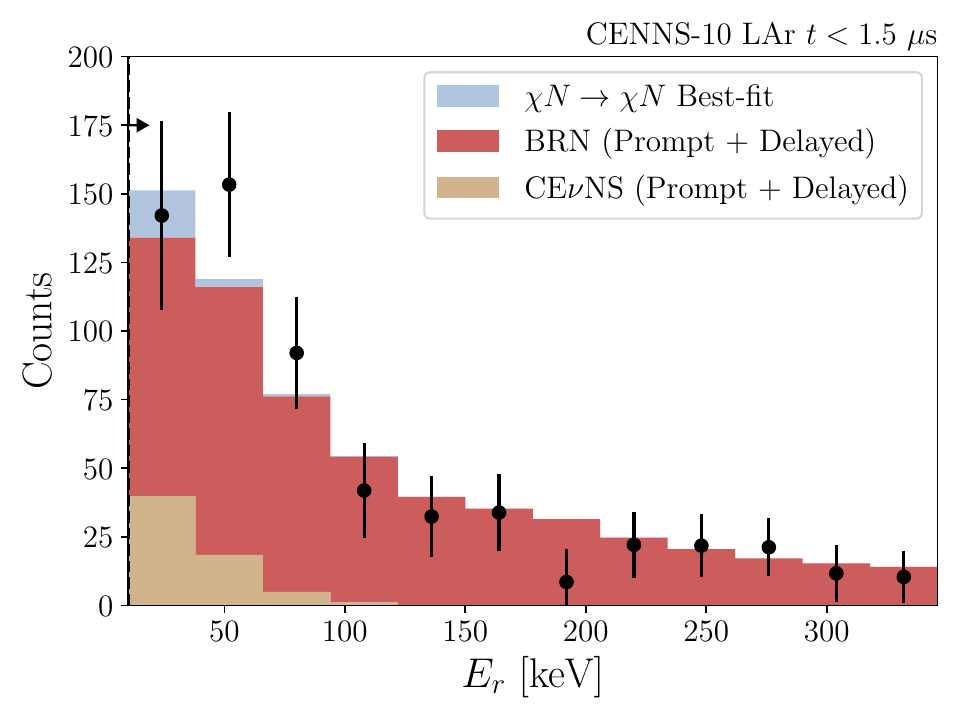}
    \caption{
    Top: 1$\sigma$-credible regions on $Y\equiv \epsilon^2 \alpha_D (\frac{m_\chi}{m_X})^4$ with $m_V/m_\chi = 3$ and $\alpha_D = 0.5$, after fitting to the COHERENT CsI data set (light blue) and the LAr data set (light green) with our energy and timing cuts in (\ref{eq:energycut}--\ref{eq:timecut}) applied.
    Best-fit recoil energy spectra with a timing cut of $t<1.5~\mu$s applied to the CsI (bottom-left) and the LAr (bottom-right) data.
    The predicted prompt-neutrino and delayed-neutrino events, the AC beam-on backgrounds measured and reported by COHERENT, and the dark matter signal with MLE parameters $(Y,m_V)=(4.5\times 10^{-9},288~{\rm MeV})$ (CsI) and $(1.9\times 10^{-11}, 36.4~{\rm MeV})$ (LAr) are stacked together.
    The COHERENT data (coincidence and anti-coincidence beam-on data combined) are shown with statistical and systematic errors added in quadrature.
    The fits to the CsI and LAr data sets with cuts are respectively conducted with the data points lying in $14~{\rm keV}<E_r<26~{\rm keV}$ and in $E_r > 10~{\rm keV}$ which are indicated by the arrows and the vertical dashed lines and defined by our energy cuts in \eqref{eq:energycut} and \eqref{eq:energyupcut}.
    }
    \label{fig:multinest}
\end{figure}

We display 1$\sigma$ best-fit regions to explain the excess by the light-blue bands in the top panel of Figure~\ref{fig:multinest}. 
The parameter space consistent with the excess lies mostly in the region $m_V > 100$ MeV at the $1\sigma$-level, since this region is where the recoil spectrum takes on a larger tail in the energy range of the cut-window, while at lighter masses, the produced spectrum is likely too steeply falling to accommodate the excess events in this window.
A few representative pairs of $(m_V,Y)$ are presented in Table~\ref{tab:best_fit}, and the corresponding $\epsilon$ values are consistent with the existing limits.

\begin{table}[t]
    \centering
    \begin{tabular}{c|c c c c}
    \hline \hline
         $m_V$ [MeV] & 75 & 100 & 150 & 200  \\
         \hline
        $Y$ (CsI) & -  & - & $6.6\times10^{-10}$ & $1.2\times10^{-9}$ \\
        \hline
        $Y$ (LAr) & $9.1\times10^{-11}$  & $2.1\times10^{-10}$ & $9.8\times10^{-10}$ & $1.8\times10^{-9}$ \\
        \hline \hline
    \end{tabular}
    \caption{Best-fit values of $Y$ for several choices of $m_V$ as derived from the results shown in the left panel of Figure~\ref{fig:multinest}. 
    For these values, $m_V/m_\chi$ and $\alpha_D$ are set to be 3 and 0.5, respectively.
    }
    \label{tab:best_fit}
\end{table}

The bottom-left panel of Figure~\ref{fig:multinest} shows the corresponding $E_r$ spectrum at the CsI detector with a timing cut of $t<1.5~\mu$s applied.
The dark matter prediction for the maximum likelihood estimate (MLE) from the likelihood scan $(Y,m_V)=(4.5\times 10^{-9},288~{\rm MeV})$ is shown. 
While Table~\ref{tab:best_fit} gives a set of example best-fit parameter points where dark photons are produced relativistically, this plot demonstrates that this choice for $m_V$ resulting in dark photon production (nearly) at rest explains the excess equally well, as suggested by the top panel of Figure~\ref{fig:multinest}. 
The best-fit dark matter signal spectrum (blue) is stacked together with the AC beam-on backgrounds (red), the predicted delayed-neutrino (brown) and prompt-neutrino (green) events. 
The COHERENT data points are marked by the black dots with statistical and systematic errors added in quadrature. 
The fit to the data set with cuts is conducted with the data points lying in our signal region $14~{\rm keV}<E_r<26~{\rm keV}$ which is defined by our energy cuts in \eqref{eq:energycut} and \eqref{eq:energyupcut}. 
This plot visualizes not only the mild excess in the signal region but the explanation of dark matter events. 

As a final remark, one may attempt the NSI interpretation as an alternative hypothesis to explain both the excess emerging after applying the cuts and the full data set. 
It was demonstrated in Ref.~\cite{Dutta:2019nbn} that the NSI hypothesis with a non-zero coupling in the electron neutrino sector does not explain the data set with the cuts and the data set without the cuts simultaneously, in particular, showing a poor fit for the excess in the prompt timing bins (i.e., $t<1.5~\mu$s). 
A non-zero coupling in the muon neutrino sector is even more disfavored since it affects the delayed neutrino events as well as the prompt neutrino events.

\subsection{Excess and dark matter interpretations - LAr Data \label{sec:explain_lar}}
The COHERENT Collaboration reported the first observation of CE$\nu$NS in liquid argon~\cite{Akimov:2020pdx} using the CENNS-10 detector with the data collected from $13.6 \times 10^{22}$ POT at the SNS.
In the CENNS-10 data release~\cite{Akimov:2020data}, the data is binned in three dimensions; like the CsI data, recoil energy in keVee and trigger time in $\mu$s make up two of the dimensions, but there is a third variable, F90, corresponding to the light yield fraction in the first 90 ns of the photo-multiplier tube response.
We use the probability distribution functions (PDFs) provided in the data release for the expected CE$\nu$NS rates as well as the BRN (prompt and delayed) and SS backgrounds, binned in $E_r$, $t$, and F90.
For the uncertainties on these PDFs, there are both normalization uncertainties and systematic or shape uncertainties.
Normalization uncertainties of 13\%, 30\%, 100\%, and 0.797\% are used for the CE$\nu$NS, prompt BRN, delayed BRN, and SS PDFs, respectively.

For the systematic uncertainties, alternate PDFs are provided in the data release to encapsulate $\pm 1\sigma$ systematic variations of the BRN and CE$\nu$NS PDFs.
For each one of the systematics (five in total), we use the alternate PDF shapes to construct continuous transformations of the default PDFs away from their mean values.
We do this in the following way: we represent the default, $+1\sigma$, and $-1\sigma$ PDFs as vectors of their bin content, $\vec{y}_0$, $\vec{y}_+$, and $\vec{y}_-$, respectively.
Now define the differences between the $\pm 1 \sigma$ alternate PDFs and the default PDFs;
\begin{equation}
    \Delta \vec{y}_\pm \equiv \vec{y}_\pm - \vec{y}_0\,.
\end{equation}
The continuous deformation of the default PDFs by the systematic shape uncertainties can then be controlled by a parameter $u \sim U(0,1)$ through the construction below:
\begin{align}
    \eta_+ (u) &\equiv \Theta\bigg(u-\frac{1}{2}\bigg) \cdot \sqrt{2}\bar{\sigma} \text{erf}^{-1} (2u-1)\,, \nonumber \\
    \eta_- (u) &\equiv \bigg[1-\Theta\bigg(u-\frac{1}{2}\bigg)\bigg] \cdot \sqrt{2}\bar{\sigma} \text{erf}^{-1} (2(1-u)-1)\,,
\label{eq:etas}
\end{align}
where ${\rm erf}(x)$ represents the error function and where the default PDFs are transformed as
\begin{equation}
    \vec{y}_0 \to \vec{y}_0 + \eta_+(u)\Delta\vec{y}_+ + \eta_-(u)\Delta\vec{y}_- \,.
\end{equation}
In Eq.~\eqref{eq:etas}, $\Theta$ is the Heaviside step function and $\bar{\sigma} \approx 2.1041$ is a width parameter that is contrived such that $\eta_+(0.6827\dots) = \eta_-(1-0.6827\dots) = 1$. This guarantees that the $+1 \sigma$ ($u=0.6827\dots$) and $-1\sigma$ ($u=0.3173\dots$) deviations of $u$ map $\vec{y}_0 \to \vec{y}_\pm$.

When we carry out the fit using the CENNS-10 data, a preliminary fit using the full timing, energy, and F90 ranges is performed with all systematic and statistical parameters floating freely in the fit, in addition to the two model parameters ($\epsilon$ and $m_V$ in the dark photon model).
We use this preliminary fit to constrain the systematic and statistical parameters, fix them, and perform a secondary fit within the LAr cut window ($t \in [0.0, 1.5]$s, $E_r \in [10, 40]$ keV).

We also exhibit 1$\sigma$ credible-regions consistent with the LAr after-cut data by the light-green band in the top panel of Figure~\ref{fig:multinest}.
We see a preferred band of $(m_V, Y)$ combinations, constituting a (roughly) flat region for $m_V \lesssim 100$ MeV where the scattering cross section given in Eq.~\eqref{eq:nuclear_xs} becomes approximately $m_V$-independent because of $m_N \gg m_V$, and a sloped region for $m_V \gtrsim 100$ MeV where the cross section starts being sensitive to $m_V$. Unlike the CsI fit, parameter space below $m_V < 100$ MeV remains consistent with the LAr data since the coarse binning of the data are likely less sensitive to changes in the spectral shape as a function of $m_V$, leaving the whole range of tested masses viable to accommodate the fit.
Below this preferred light-green band, there are scattered small credible ``islands" which express mild consistency with the data down to the null limit $Y \to 0$.
However, the overlapping credible regions of parameter space within $140~{\rm MeV}\lesssim m_V \lesssim 500~{\rm MeV}$ accommodate both the CsI and LAr data after applying our timing and energy cuts.

The bottom-right panel of Figure~\ref{fig:multinest} shows the corresponding best-fit $E_r$ spectrum with a timing cut of $t<1.5~\mu$s applied to the LAr data. The energy cut $E_r > 10$ keVnr is shown by the dotted line.
The dark matter prediction for MLE from the likelihood scan $(Y,m_V)=(1.9\times 10^{-11}, 36.4~{\rm MeV})$ is shown.
The best-fit dark matter signal spectrum (blue) is stacked together with the BRN background (red) and the predicted CE$\nu$NS events (brown). 
The COHERENT data points are marked by the black dots with statistical and systematic errors added in quadrature. 
The fit to the data set with cuts is conducted with the data points lying in our signal region $E_r > 10~{\rm keV}$ which is defined by our energy cuts in \eqref{eq:energycut}.

\subsection{Constraining parameter space \label{sec:constrain}}

As mentioned in the preceding section, the ``excess'' may be explained by an unidentified background or a systematic uncertainty on the observed steady-state background.
It may also disappear as more statistics are taken into account.
In this case, our proposed analysis strategy can improve the experimental sensitivities to the models discussed in Section~\ref{sec:models}, as it allows for eliminating a large portion of backgrounds.
Likewise, for our benchmark detectors other than the COHERENT CsI detector, we can study the expected experimental sensitivity reaches together with the cuts defined in Section~\ref{sec:analysis}.
In order to evaluate future sensitivities at COHERENT, CCM, and JSNS$^2$ to dark matter signals, we again perform a likelihood analysis using simulation data at each experiment for nominal choices of the expected exposure.
Given that data and information about backgrounds are available for COHERENT CsI and LAr, we treat those likelihood analyses differently than those for CCM and JSNS$^2$.

\begin{itemize}
\item {\bf COHERENT}:
We simulate a scenario where the CsI excess vanishes with more exposure.
To do this, we use simulated data based on the null hypothesis as our ``observed" data with a 3-year run period, combining both LAr and CsI data. 
The sizes of the neutron distribution $R_n$ are set to be 4.7~fm and 4.1~fm~\cite{Cadeddu:2020lky} for CsI and Ar, respectively.
Backgrounds as well as prompt and delayed neutrino rates are scaled accordingly, but systematic uncertainties are kept the same.
We then define the following test statistic:
\begin{equation}
\chi^2 \propto \sum_\text{CsI,LAr}\sum_{(t,E_r)} \dfrac{(N(\vec\theta) - N_\nu - N_\text{bg})^2}{(N_\nu + N_\text{bg})(1 + \sigma^2 (N_\nu + N_\text{bg}))},
\end{equation}
for $N(\vec\theta) = N_\chi (\vec\theta) + N_\nu + N_\text{bg}$, which is based on a null hypothesis event rate $N_0 = N_\nu + N_{\rm bg}$ in each $(t,E_r)$ bin. We adopt the simplifying assumption, for the sake of estimating future sensitivity, of flat background distributions whose total rates are given in Refs.~\cite{Akimov:2017ade,Akimov:2020pdx} and scaled linearly to account for a 3-year exposure.
The total systematic uncertainty $\sigma$ is again assumed to remain at 28\% for CsI~\cite{Akimov:2017ade} and 8.5\% for LAr~\cite{Akimov:2020pdx}.
In order to evaluate the sensitivity to the model parameters $\vec{\theta}$, we find the contour in parameter space of 90\% C.L.. 

\item {\bf CCM and JSNS$^2$}:
We use the same test statistic for 3-year run periods, but we do not include SS, BRN, or other background rates as they are unknown at this time.
We also do not include the treatment of systematic uncertainties for the same reason, i.e., $\sigma = 0$ for CCM and JSNS$^2$.
Again, the size of the neutron distribution $R_n$ for Ar is set to be 4.1~fm.
\begin{equation}
\chi^2 \propto \sum_{(t,E_r)} \dfrac{(N(\vec\theta) - N_\nu)^2}{N_\nu}\,. \end{equation}
\end{itemize}

We now discuss the parameter space for two different scenarios that we specified at the beginning of this section, i.e., single-mediator  and double-mediator scenarios. 
For both possibilities, we take dark photon as our mediator appearing in the dark matter scattering process for illustration. 
Therefore, both quark- and lepton-related production channels including P1 through P4 come into play in the single-mediator scenario, while signal detection via nucleus scattering D1 (for COHERENT and CCM) and electron scattering D2 (for JSNS$^2$) is available in both scenarios.
In the double-mediator scenario, we will choose baryon number-gauged dark gauge boson for production of dark matter as a concrete example, while arguing that the dark photon nevertheless governs the scattering signal.
The main purpose of showing two different types of scenarios is due to the fact that in any realistic model, there can be more than one mediators associated with multiple gauge bosons, scalars, etc.

\medskip

\begin{figure}[t]
    \centering
    \includegraphics[width=15cm]{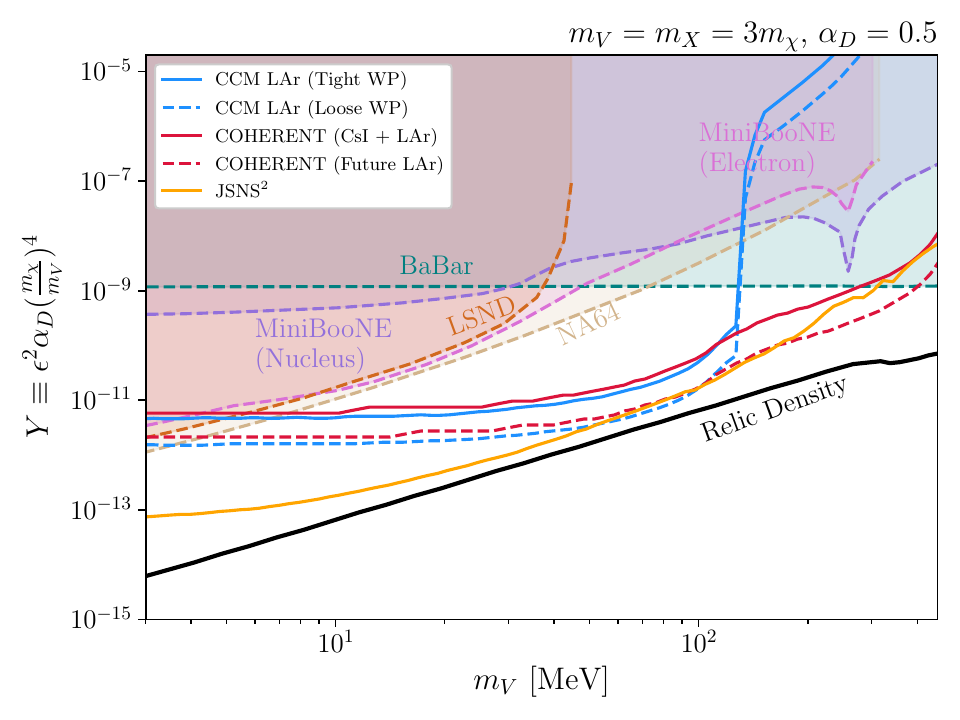}
    \caption{90\% C.L. projected experimental sensitivity to the model couplings and mediator masses in the single-mediator scenario ($X = V$) for our benchmark detectors.
    We take a dark photon $A'$ as our mediator, and conventionally plot the sensitivities on $Y\equiv \epsilon^2 \alpha_D (\frac{m_\chi}{m_V})^4$ where $\alpha_D = (\kappa^V_D)^2 / (4\pi)=0.5$ and $m_V/m_\chi=3$.
    The relevant existing limits from BaBar~\cite{Lees:2017lec}, LSND~\cite{deNiverville:2011it}, MiniBooNE~\cite{Aguilar-Arevalo:2018wea}, and NA64~\cite{NA64:2019imj} are shown by the shaded regions. 
    The parameter sets that are consistent with the observed dark matter relic abundance are shown by the black solid line.
    } 
    \label{fig:limits_single_mediator}
\end{figure}

\noindent {\bf Single-mediator scenario}: 
In Figure~\ref{fig:limits_single_mediator}, we exhibit the expected 90\% C.L. sensitivity reaches for COHERENT with CsI and LAr (current and future), CCM, and JSNS$^2$ in the context of the single-mediator scenario, with the mass ratio of dark matter to the mediator fixed to be 3. 
We plot $Y=\epsilon^2\alpha_D(m_\chi/m_V)^4$ versus the mediator mass $m_V$, using our energy and timing cuts from \eqref{eq:energycut} and \eqref{eq:timecut}.
We find that P1 (i.e., meson decays described in section~\ref{sec:production}) make a dominant contribution to the determination of sensitivity reaches, while the others are subleading.
The relevant existing limits from BaBar~\cite{Lees:2017lec}, LSND~\cite{deNiverville:2011it}, MiniBooNE~\cite{Aguilar-Arevalo:2018wea}, and NA64~\cite{NA64:2019imj} are also shown for comparison, and the excluded regions are shaded by respective colors. 
We also show the thermal relic abundance line (black solid) consistent with the observed abundance in Figure~\ref{fig:limits_single_mediator} where the mediator couples to electron and quarks.
The relic density calculation presented here is obtained semi-analytically including all available channels ($\chi\bar{\chi}\to e^+e^-, \mu^+\mu^-, \pi^+\pi^-$, etc.)
in increasing $m_V$, and the result agrees with the \texttt{MicrOMEGAs}~\cite{Belanger:2018ccd} calculation for the same parameter space.  

For the COHERENT CsI and LAr, we assume no excess in calculating the limits, while the existing data may have a mild excess as discussed previously.
One should notice that our projected COHERENT limit is smoothly extended beyond $m_V\sim 130$~MeV as we separately include the additional $\eta$ production via nuclear motion~\cite{Akimov:2019xdj,Cassing:1991jr}.
We find that the COHERENT limit based on the ongoing CsI and LAr detectors can cover parameter regions unexplored by other existing searches for $m_V\sim 5-200$~MeV.
We also show the COHERENT future LAr line (red dashed) for 610 kg fiducial mass for 3 years using our energy and timing selections. 
The limit suggests that our proposed strategy allow the COHERENT LAr to explore a wider range of parameter space, as compared to the projected limit in the experimental analysis shown in Ref.~\cite{Akimov:2019xdj}. 
In the experimental analysis, the prompt timing window $t\leq 1.5~\mu$s is used, but no energy cut is employed to remove the prompt neutrino.
A side-band measurement is made on the delayed neutrino ($\nu_e$ and $\bar{\nu}_\mu$) to determine the shape of prompt neutrino ($\nu_\mu$) in order to remove it from the dark matter analysis.  
This side-band analysis depends on the the assumption that different neutrino flavors only possess SM interactions.   
On the other hand, if we use the energy cut on the prompt window, we reject the neutrino events even if they possess NSIs for different flavors. 
Hence, the energy cut along with prompt-window selections can probe dark matter by vetoing neutrinos even with the NSI interactions.  
Furthermore, with the energy cut, stronger bounds are allowed for $m_V\gtrsim 10$~MeV in our analysis compared to those in the Ref.~\cite{Akimov:2019xdj}.
The approach therein is basically to determine the prompt neutrino distribution up to statistical and systematic errors by a side-band analysis. 
It allows for constrained systematics, but the prompt neutrino events remain not eliminated unlike our kinematic cuts. Therefore, as the mediator mass increases, the number of dark matter events relative to the neutrino and background event rates gets smaller so that the resultant limit becomes weaker rapidly.

For CCM with a LAr detector, we show the projected limits (blue dashed and solid) with both tight and loose timing cuts defined in \eqref{eq:timecut}.
The tight cut is suggested by the CCM Collaboration~\cite{privateCCM}, whereas the loose cut is determined by our cut optimization.
The improvement with the loose cuts is about 50\% for the limit on the coupling compared to the tight-cut case. 
Due to the larger detector mass (7 tons fiducial), the reach is expected to be better than the COHERENT future with 610~kg of LAr fiducial mass. Since the energy of the proton beam for CCM is 800~MeV, there are few energetic $e^\pm$ and $\eta$ mesons created, and as a result the maximum reach over unconstrained parameter space for $m_V$ is about 300~MeV. 

For JSNS$^2$ with a Gd-LS detector, we utilize the first pulse out of the two consecutive pulses as discussed in Section~\ref{sec:analysis} in order to perform the analysis with a lower neutrino flux. 
The projected limit line (orange) is expected to be better than the other experiments due to a large volume (17 tons fiducial) of the detector.
We see that the JSNS$^2$ sensitivity line shows a curvature different from the other experiments, i.e., smoothly rising trend versus flat trend in increasing $m_V$.
The reason is that the mass differences between the mediator and the target particle off which dark matter scatters are differently hierarchical~\cite{Kim:2020ipj}. 
In more detail, the target electron for JSNS$^2$ is lighter than the mediator, so the scattering cross section decreases by $\sim 1/m_V^4$ suggested by Eq.~\eqref{eq:electron_xs}. 
By contrast, the target nucleus for COHERENT and CCM is much heavier than the mediator, and thus the scattering cross section is essentially governed by the mass of target nucleus, i.e., roughly constant (for $m_V \lesssim 100$~MeV), as suggested by Eq.~\eqref{eq:nuclear_xs}. 
Speaking of the reach for $m_V$, for COHERENT and JSNS$^2$ the coverage is extended further up to $\sim 700$~MeV due to the production of dark gauge bosons from heavier mesons (e.g., $\eta$) which can decay into heavier $V$ bosons.

It is instructive to qualitatively understand why the proposed search strategy allows our benchmark experiments, in particular, JSNS$^2$ to have significantly improved sensitivity prospects over other experiments with similar experimental specifications, e.g., LSND and MiniBooNE. 
Compared to LSND, JSNS$^2$ is capable of producing $\sim10$ times more $\pi^0$ per POT due to a higher beam energy, receiving $\sim1.6$ times more signal flux per unit area due to its smaller proximity to the detector, and significantly reducing the delayed neutrino backgrounds by an order of magnitude through an application of the timing cut (see the bottom panels of Figure~\ref{fig:spectra}). 
On the other hand, in the LSND sensitivity estimate~\cite{deNiverville:2011it}, a 19\% efficiency was taken while we assume a 100\% efficiency. Moreover, the recoil energy is restricted to 18~MeV through 50~MeV in the LSND analysis, while a lot of signal events beyond 50 MeV is expected (see the bottom panels of Figure~\ref{fig:spectra}), hence is included in our analysis. 
The MiniBooNE limits~\cite{Aguilar-Arevalo:2018wea} are based on the POTs smaller by $\sim2$ orders of magnitude than the POTs of JSNS$^2$ that we considered, and the distance between the detector and the dump at MiniBooNE is about $\sim500$ meters. 
Considering all these factors together, our estimates along with the proposed timing and energy cuts allow JSNS$^2$ to improve the existing limits from LSND and MiniBooNE significantly.

\medskip

\noindent {\bf Double-mediator scenario}:
Moving onto the double-mediator scenario, we consider two mediators with different couplings to SM fermions ($\kappa^f_V\neq \kappa^f_X$) and to dark matter ($\kappa^D_V\neq \kappa^D_X$).
It is then possible to have different cases based on various combinations of the couplings. 
As mentioned previously, we choose a dark photon $A'$ as $V$ (mediator for dark matter detection) and a baryon number-gauged dark gauge boson $B$ as $X$ (mediator for dark matter production) solely for illustration.\footnote{A similar argument can go through if one replaces $B$ by $T_{3R}$ model 2.} 
Therefore, $\kappa_f^X$ and $\kappa_f^V$ are identified as $g_B$ and $\epsilon e$, respectively (see also Table~\ref{tab:modelsummary}).
While the search under consideration is sensitive to effective coupling $\kappa_{\rm eff}$ in Eq.~\eqref{eq:effcouple}, we fix $\kappa_f^X$ to be $2 \times 10^{-3}$, which is consistent with the current experimental constraints~\cite{Artamonov:2009sz},\footnote{UV-completed anomaly-free U(1)$_B$ models may suffer from more stringent bounds~\cite{Dror:2017ehi}.} and $\alpha_D=(\kappa_D^V)^2/(4\pi)$ to be 0.5 for simplicity. 
We further take $m_X=75$~MeV and $m_\chi=2$~MeV, so hadronic decays of $X$ are kinematically forbidden while $X\to e^+e^-$ may arise via loop-suppressed couplings. 
Therefore, $X\to \chi\bar{\chi}$ can dominate unless $\kappa_D^X$ is too small. 
We set $\kappa_D^X$ to be $\sim 10^{-7}$ which allows for an almost full flux of dark matter reaching the detector within the prompt timing window, as discussed in Section~\ref{sec:timing}. 

One may ask whether these parameter choices guarantee that dark matter scatters via an exchange of mediator $V$ not $X$. 
For JSNS$^2$, this is not an issue because the associated detector is in favor of dark matter that can interact with electrons. 
For COHERENT and CCM, this works if $\kappa_f^X\kappa_D^X < \kappa_f^V \kappa_D^V$.
The mediator mass parameters $m_X$ and $m_V$ are largely irrelevant because the factor from the mediator propagator is dictated by the nucleus mass [see Eq.~\eqref{eq:nuclear_xs}]. 
We will see shortly that this condition is satisfied over the regions of parameter space that we are exploring. 

\begin{figure}[t]
    \centering
    \includegraphics[width=15cm]{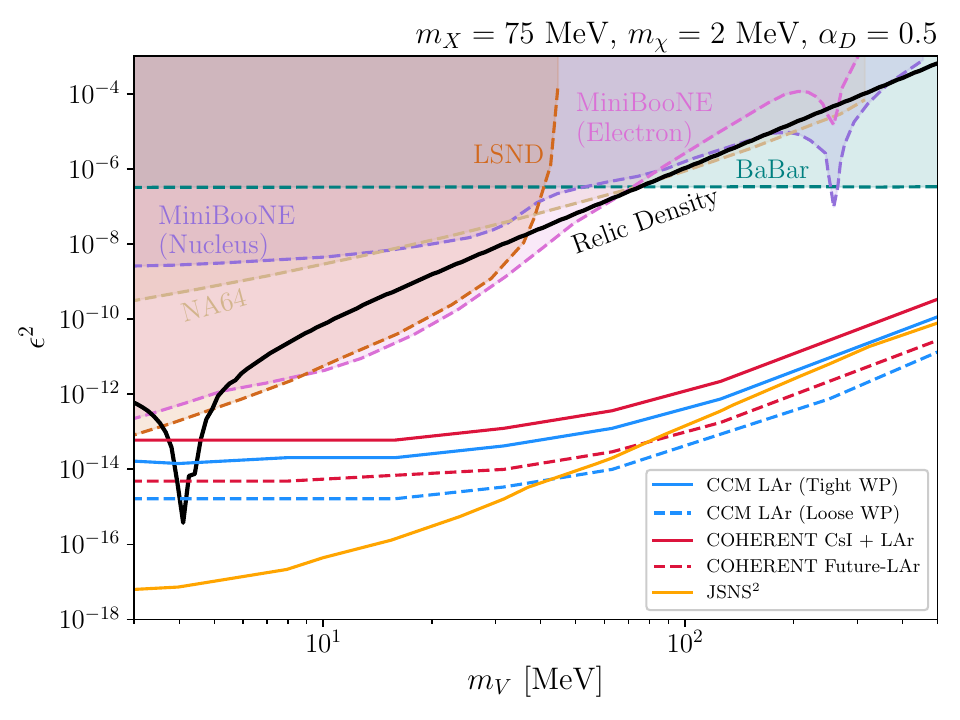}
    \caption{90\% C.L. projected experimental sensitivity to the model couplings and mediator masses in the double-mediator scenario. $X$ and $V$ are taken to be a baryon number-gauged dark gauge boson $B$ (i.e., $\kappa_f^X=g_B$) and a dark photon $A'$ (i.e., $\kappa_f^V=\epsilon e$), respectively, for illustration. 
    The masses of $X$ and dark matter are fixed to $m_X=75$~MeV and $m_\chi=2$~MeV. The coupling $\kappa_f^X$ is set to be $2\times 10^{-3}$, which is consistent with the current constraints~\cite{Artamonov:2009sz}, while we take $\kappa_D^X=10^{-7}$ and $\alpha_D \equiv (\kappa_D^V)^2 / (4\pi)=0.5$. 
    We plot the sensitivities on $\epsilon^2$, while the existing limits are recast for this scenario.
    }
    \label{fig:limits_double_mediator}
\end{figure}

Figure~\ref{fig:limits_double_mediator} displays the 90\% C.L. sensitivity reaches of our benchmark detectors in the scenario specified thus far, in the plane of $m_V$ against $\epsilon^2(=(\kappa_f^V)^2/e^2)$.
The reach for $m_V$ 
is not limited like before, since $m_X$ 
is fixed. 
The exclusion limit lines are improved  by $\sim 4$ orders of magnitude (for $m_V \lsim 100$~MeV), compared to $\epsilon^2$ calculated from $Y$ in Figure~\ref{fig:limits_single_mediator}.
All our benchmark experiments are essentially friendly to baryon-involving dark matter production channels.
Relatively less constrained baryo-philic mediator $X$ allows for copious production of dark matter, and as a consequence it becomes possible to explore the regions of smaller $\epsilon$ values. 
The CCM limit at Loose WP reaches down to $\sim 5\times 10^{-15}$, and we find that $\kappa_f^X\kappa_D^X$ is about three orders of magnitude smaller than $\kappa_f^V\kappa_D^V$, thereby satisfying the aforementioned condition and letting the dark matter scatter off a nucleus dominantly via an exchange of $V$.
The existing limits are recast for this double-mediator scenario. 
Since BaBar and NA64 set their limits based on the model assumption that the dark photon emitted from an electron beam decays invisibly, i.e., $e^+e^- \to \gamma A'(\to {\rm invisible})$ for BaBar and $e^-Z \to e^-ZA'(\to {\rm invisible})$ for NA64, their limits are simply raised from those in Figure~\ref{fig:limits_single_mediator} by a factor of $\frac{1}{\alpha_D}\left(\frac{m_V}{m_\chi}\right)^4$. 
However, MiniBooNE and LSND basically assumed the dark gauge boson production from meson decays, so their limits can be reinterpreted in the context of the specific scenario that we consider here,\footnote{We further assume that dark matter production in MiniBooNE and LSND is dominated by the $\pi^0$ decay over the others.} thereby setting more stringent limits than those in Figure~\ref{fig:limits_single_mediator}. 
We will discuss how to recast the limits from MiniBooNE and LSND in more detail in the next section. 

For the dark matter relic abundance, we remark that the domination of the $V$ boson also would continue in the relic abundance calculation under the assumption that dark matter $\chi$ is thermally frozen out in the early universe. 
Given the fact that the chosen $X$ boson is baryo-philic and $\chi$ is set to be 2~MeV, there are no tree-level channels to pair-annihilate $\chi$ to the SM (hadronic) final state around the freeze-out time, while loop-induced couplings can lead to leptonic annihilation channels, e.g., into an electron-positron pair.
By contrast, annihilation channels via the $V$ boson are available even at the tree level, e.g., $\chi\bar{\chi}\to V^* \to e^+e^-$.\footnote{For $m_V < m_\chi$, another channel, $\chi\bar{\chi}\to VV$ may open and change details of the $\chi$ freeze-out.}
Therefore, if $\dfrac{(\kappa_f^V \kappa_D^V)^2}{m_V^4}>\dfrac{(c_{\rm loop} \kappa_f^X \kappa_D^X)^2}{m_X^4}$, the resulting relic abundance can be (mainly) dictated by the $V$ boson. 
Our choices for $\kappa_f^X$ and $\kappa_D^X$ multiplied by the loop-induced factor $c_{\rm loop}$ are significantly smaller than the product $\kappa_f^V\kappa_D^V$, so that the $X$ boson essentially does not affect the relic abundance within the range of $m_V$ shown in Figure~\ref{fig:limits_double_mediator}. 
We display the corresponding relic density curve again by the black solid line, assuming that $\chi$ is the cosmological dark matter. 
The dip around $m_V=4$~MeV originates from the ($s$-channel) resonance annihilation at $m_V \approx 2m_\chi$.
We see that even the existing COHERENT data (${\rm CsI}+{\rm LAr}$) would allow us to explore the regions of parameter space below the relic density line for the given set of values for $\alpha_D$, $\kappa_f^X$, and $\kappa_D^X$.

\medskip

\subsection{Interpretations in other dark matter scenarios \label{sec:recast}}

We have discussed so far the dark matter signal sensitivity of our benchmark experiments in the context of a certain specific dark matter scenario (mostly a dark photon scenario) purely for illustration. 
As listed in Table~\ref{tab:modelsummary}, there are a variety of dark matter scenarios to which the benchmark experiments are sensitive, so we discuss ways of not only interpreting our results in the context of various models in this section but also determining appropriate existing limits and recasting them properly. 
This exercise depends on the underlying model assumptions.
We discuss them in the single-mediator scenario followed by the double-mediator scenario.  

The expected number of dark matter events in the detector $N_{\chi}$ is proportional to the dark matter flux $\Phi_\chi$ and the scattering cross section between dark matter and the target particle $T$ (e.g., $T=N, e^-$): $N_\chi \propto \Phi_\chi \sigma_{\chi T}$. 
Analyses are usually done for individual  detection channel; for example, MiniBooNE reported their results in the electron channel and the nucleus channel. 
Once the channel is fixed, the overall cross section $\sigma_{\chi T}$ becomes proportional to the associated coupling constant(s) and the experiment becomes exclusively sensitive to those coupling constant(s). 
However, $\Phi_\chi$ is inclusive; for our benchmark experiments, $\Phi_\chi = \Phi_{\chi,{\rm P1}}+\Phi_{\chi,{\rm P2}}+\Phi_{\chi,{\rm P3}}+\Phi_{\chi,{\rm P4}}+\cdots$ up to other negligible contributions. 
Depending on the choice of dark gague boson $X$, a subset of the production channels get suppressed; for example, the flux via the $e^\pm$-induced cascade $\Phi_{\chi,{\rm P4}}$ is suppressed for baryo-philic scenarios. 
Furthermore, even among the allowed channels the size of the dark matter flux in one channel relative to those in the others may change depending on the type of dark gauge boson scenario we choose. 
This complication can be avoided if $\Phi_\chi$ is dominated by a single channel. 
We find that in our benchmark experiments a majority of dark matter is produced via the $\pi^0$ decay, so it is a reasonably good approximation to take $\Phi_\chi \approx \Phi_{\chi,{\rm P1}}$.
For the electro-philic $X$, the meson-involved channels (i.e., P1, P2, and P3) are suppressed, so $\Phi_\chi$ is then approximated to $\Phi_{\chi,{\rm P4}}$. 

In the single-mediator case ($X=V$, $\kappa_f^X=\kappa_f^V$, and $\kappa_D^X=\kappa_D^V$), for a given $m_V$ the translation rule between the $\epsilon$ value in Figure~\ref{fig:limits_single_mediator} and $\kappa_f^V$ can be obtained by equating the products of the dominant dark matter flux and the coupling associated with the dark matter-nucleus scattering. 
For example, a baryo-philic scenario (e.g., $B$ and $T_{3R}$ model 2) takes the $\pi^0$ channel as the dominant dark matter production channel like the dark-photon scenario, so we find
\begin{equation}
    \frac{g_B^2}{\epsilon^2e^2}=\frac{Z}{A}\frac{\kappa_D^{A'}}{\kappa_D^B} ~~\hbox{for $B$}\quad \hbox{and}\quad
    \frac{g_{T_{3R}}^2}{\epsilon^2e^2}=\frac{Z}{2(A-2Z)}\frac{\kappa_D^{A'}}{\kappa_D^{T_{3R}}} ~~\hbox{for $T_{3R}$}\,,
\end{equation}
which are more relevant to COHERENT and CCM because the JSNS$^2$ detector is less sensitive to dark matter interacting with quarks. 
The COHERENT and CCM experiments depend mostly on the quark couplings for the dark matter production, while the detection of dark matter depends entirely on quark couplings of the associated dark gauge boson because these detectors are optimized to small energy deposits.
The direct electron coupling for these experiments appears in the estimation of the $e^\pm$-induced bremsstrahlung production of the dark gauge boson.
Therefore, if the mediator participating in detection does not have tree-level quark coupling, these experiments would be less sensitive to the associated dark matter models. 
On the other hand, the dark matter detection at JSNS$^2$ depends entirely on the electron coupling because the energy threshold of the detector disfavors nuclear recoil, whereas dark matter can be created via both quark and electron couplings.
Therefore, if an underlying dark matter model were ``electro-philic'', JSNS$^2$ would be more advantageous than the other two in the search for the dark matter signal.\footnote{A lepto-philic model $L_\mu-L_\tau$ would not be tested at JSNS$^2$ as efficiently as other lepto-philic models with direct coupling to electrons.}
For dark gauge bosons carrying both quark and electron couplings like dark photon, all three experiments are capable of producing and detecting dark matter. 
In summary, the COHERENT, CCM, and JSNS$^2$ experiments would provide complimentary information on couplings for a dark sector model, and their sensitivity to various dark sector models is summarized in Table~\ref{tab:single-sense-summary}. 
For the $T_{3R}$ model, the sensitivities for model 2 are shown in the parentheses only for the cases which differ from model 1.

\begin{table}[t]
    \centering
    \resizebox{\columnwidth}{!}{%
    \begin{tabular}{c|c c c c c c c}
    \hline \hline
    Type of mediator $V$ & $A'$ & $B$ & $L$ & $B-L$ & $L_e-L_\mu$ & $L_e-L_\tau$ & $T_{3R}$ [model 1(2)]  \\
    \hline
    COHERENT & \cmark & \cmark & \xmark & \cmark & \xmark & \xmark & \cmark  \\
    CCM  & \cmark & \cmark & \xmark & \cmark & \xmark & \xmark & \cmark  \\
    JSNS$^2$  & \cmark & \xmark & \cmark & \cmark & \cmark & \cmark & \cmark (\xmark)  \\
    \hline
    LSND~\cite{deNiverville:2011it}  &  \cmark & \xmark & \cmark & \cmark & \cmark & \cmark & \cmark (\xmark)   \\
    MiniBooNE--Electron~\cite{Aguilar-Arevalo:2018wea}  &  \cmark & \xmark & \cmark & \cmark & \cmark & \cmark & \cmark(\xmark)   \\
    MiniBooNE--Nucleus~\cite{Aguilar-Arevalo:2018wea}  &  \cmark & \cmark & \xmark & \cmark & \xmark & \xmark & \cmark   \\
    BaBar~\cite{Lees:2017lec}  & \cmark & \xmark & \cmark & \cmark & \cmark & \cmark & \cmark   \\
    NA64~\cite{NA64:2019imj}  & \cmark & \xmark & \cmark & \cmark & \cmark & \cmark & \cmark  \\
    \hline \hline
    \end{tabular}
    }
    \caption{The sensitivity of our benchmark experiments, COHERENT, CCM, and JSNS$^2$, and existing relevant experiments, BaBar, LSND, MiniBooNE, and NA64, to the dark sector models listed in Table~\ref{tab:modelsummary}, whose mediator governs the scattering process of dark matter. 
    The ``\cmark'' (``\xmark'') symbol indicates that the experiment in the associated row is sensitive (less sensitive) to the model in the associated column through tree-level interactions.} 
    \label{tab:single-sense-summary}
\end{table}

Recasting the limits from LSND and MiniBooNE can be obtained in a similar fashion, as their search scheme also relies on the ``appearance'' of proton-beam-produced dark matter (mostly from $\pi^0$ decays) through its scattering-off electrons (LSND and MiniBooNE--Electron) or nuclei (MiniBooNE--Nucleus).  
However, as discussed earlier, BaBar and NA64 rely on the ``disappearance'' of the dark matter radiated off from an electron beam. Therefore, they are not capable of setting stronger limits for baryo-philic models (i.e., through tree-level couplings). 
In addition, they are only sensitive to $\kappa_f^V$, the resulting rescaling is simply determined by the ratio of gauge charges of electron:
\begin{equation}
    \frac{\kappa_f^V}{\epsilon e} = \frac{x_e^V}{x_e^{A'}}\,.\label{eq:recast-disappearance}
\end{equation}
The sensitivity of the above five experiments to various dark sector models is also summarized in Table~\ref{tab:single-sense-summary}.

In the double-mediator case ($X\neq V$, $\kappa_f^X\neq \kappa_f^V$, and $\kappa_D^X\neq \kappa_D^V$), the sensitivity plot is usually shown in the $m_V-\kappa_f^V$ plane as in Figure~\ref{fig:limits_double_mediator}. 
For the ``appearance'' experiments such as COHERENT, CCM, JSNS$^2$, LSND, and MiniBooNE, the detection channel should be sensitive to the model of interest, i.e., mediator $V$. 
By contrast, the two ``disappearance'' experiments here, BaBar and NA64, again can set the limits as long as $V$ is electro-philic. 
Therefore, the summary in Table~\ref{tab:single-sense-summary} goes through for mediator $V$ of a given double-mediator scenario. 

Recasting the limits from BaBar and NA64 simply follows the rule in Eq.~\eqref{eq:recast-disappearance}, whereas care must be taken to recast the limits from LSND and MiniBooNE. 
If dark gague boson $X$ has couplings to up and down quarks, dark matter production is dominantly from $\pi^0$ decays and it is sufficient to compare the products of relevant coupling constants, resulting in
\begin{equation}
    \kappa_f^V = \frac{\frac{1}{3}(\epsilon e)^2 Q_{\rm eff}^{A'} \kappa_D^{A'}}{x_{\pi^0}^X\kappa_f^X Q_{\rm eff}^V\kappa_D^V}\,,
\end{equation}
where $\epsilon$ is deduced from the limits of LSND/MiniBooNE, $\kappa_D^{A'}$ is from the assumed $\alpha_D$ of LSND/MiniBooNE, and $Q_{\rm eff}^{A'}$ is $Z$ for the MiniBooNE--Nucleus while the MiniBooNE--Electron and LSND take $\sqrt{Z}$ instead.
$1/3$ in the numerator and $x_{\pi^0}^X$ in the denominator are the effective gauge charges of $\pi^0$ with respect to $A'$ and $X$, respectively, while $Q_{\rm eff}^V$ is the effective charge associated with mediator $V$ for the nucleus scattering and is replaced by $\sqrt{Z}x_e^V$ for the electron scattering. 

One may then fix $\kappa_f^X$ to a value to explore a slice of parameter space spanned in the $m_V-\kappa_f^V$ plane, as is done in Figure~\ref{fig:limits_double_mediator}.
However, as briefly discussed in the previous section, particular care should be taken for choosing a value of $\kappa_f^X$ in order to be consistent with three main underlying assumptions in the double-mediator scenario: 
($i$) dark matter is dominantly produced through $X$, ($ii$) produced $X$ predominantly decays into a dark matter pair, and ($iii$) the dark matter scattering arises predominantly via an exchange of mediator $V$. 
Note that these assumptions are more for convenience of analyses hence developing the intuition on the experimental sensitivity to couplings possibly one at a time.
One can perform an analysis and recasting with all or part of the above assumptions relaxed, at the expense of complicating the analysis and the recasting. 
Assumption ($i$) is relevant if $V$ has also couplings to quarks, and can be readily satisfied as far as the fixed value of $\kappa_f^X$ is larger than the value of $\kappa_f^V$ near the experimental reach modulo associated gauge charges.
Assumption ($ii$) itself can be satisfied as far as $\Gamma_{X\to \chi\bar{\chi}} \gg \Gamma_{X\to f\bar{f}}$. This relation (roughly) holds for $\kappa_D^X>\kappa_f^X$, but as we have seen in the previous section, a fairly large $\kappa_D^X$ on top of a sizable $\kappa_f^X$ may be in conflict with assumption ($iii$).
Conversely, too small coupling constants would lead too long a lifetime of $X$, resulting in a substantial reduction of the dark matter flux reaching the detector.
Depending on the situation, $\kappa_D^X$ should be sensibly selected for consistency among underlying assumptions.
For baryo-philic $X$ with $m_X<2m_\pi$, the decay modes of $X$ to the SM particles can be significantly suppressed at the tree level, and the loop-induced leptonic decay channels can allow more space for reasonable choices of $\kappa_f^X$ and $\kappa_D^X$. 
Finally, the assumption ($iii$) is relevant whenever $V$ and $X$ are competing in the scattering process of dark matter.
In the case of nucleus scattering with $m_N\gg m_{X/V}$, $(\kappa_f^V \kappa_D^V)^2 > (\kappa_f^X \kappa_D^X)^2$ is needed up to $Q_{\rm eff}^{X/V}$, while in the case of electron scattering with $m_{X/V} \gg m_e$, $\frac{(\kappa_f^V \kappa_D^V)^2}{m_V^4} > \frac{(\kappa_f^X \kappa_D^X)^2}{m_X^4}$ up to gauge charges of electrons $x_e^{X/V}$. 
Again parameter choices to satisfy assumption ($iii$) may disfavor assumption ($ii$), so one should check whether all the assumptions hold consistently for a given set of parameter choices.

\section{Conclusions \label{sec:conclusions}}

Low-energy and high-intensity beam-based neutrino experiments are receiving increasing attention as an excellent venue for probing new physics not only in the neutrino sector but in the dark matter sector. 
As expected, the neutrinos are a major contaminant to dark matter searches in these experiments. 
This situation is similar to the ``neutrino floor'' in the dark matter direct detection experiments.
We proposed recoil energy and timing selections to overcome the neutrino floor which allows the ongoing and upcoming experiments to have much better sensitivity reaches to the dark matter signal, compared to the existing limits.
The cuts that we develop remove not only the neutrinos with the SM interactions but also with NSIs.

In this work, we first showed the irreducible neutrino floor associated with COHERENT (CsI and LAr) , CCM (LAr) and JSNS$^2$ (Gd-LS). 
We then determined the recoil energy and timing selections for different types of detectors for these experiments to reduce the prompt and delayed neutrino event rates. 
The cut selections vary with the detector type. 
All these stopped-pion experiments not only produce neutrinos but they are high-intensity sources for photons emerging from meson decays and bremsstrahlung. 
We utilized all these photons to produce dark gauge bosons which subsequently decay into dark matter. 
We then investigated possible searches for dark matter from the nuclear and electron recoils at the detector. 
Recoils associated with the appearance of dark matter render these experiment robust and complimentary to various beam-dump experiments where the disappearance of dark gauge boson is used to put limits. 
Further, since the JSNS$^2$ experiment uses electron couplings of the dark gauge bosons to look for the dark matter appearance, whereas the CCM and the COHERENT experiments use the quark couplings of the dark gauge bosons, we obtain complimentary information about these dark matter models from the three experiments.

The COHERENT experiment has already published results associated with CsI and LAr detectors~\cite{Akimov:2017ade,Akimov:2020pdx}, while the CCM and the JSNS$^2$ experiments are ongoing but without any public results yet.
The measurement data of the CsI and LAr detectors in the COHERENT experiment is publicly available, so we analyzed the data using our proposed event selection scheme.
We found a mild excess and demonstrated that our dark matter models can accommodate the excess~\cite{Dutta:2019nbn}.
We then investigated dark matter signal sensitivity expected in all these experiments, showing the limit plots in the plane of the mediator mass versus a coupling parameter in the single-mediator and the double-mediator scenarios.
In the single-mediator scenario, the current limit emerging from the COHERENT data appears to already improve the existing limits from NA64, MiniBooNE, BaBar, etc. for a similar parameter space. 
Our study suggests that other experiments be capable of probing wide ranges of unexplored parameter space, getting closer to the thermal relic density line. 
In the double-mediator scenario, we found that our benchmark experiments can probe even wider ranges of unexplored parameter space by allowing a baryo-philic dark gauge boson to be responsible for dark matter production. 
Furthermore, all of our benchmark experiments rely on the ``appearance'' of produced dark matter via its scattering process, so their limits can be considered more robust than the experiments relying on the ``disappearance'' of produced dark matter, e.g., BaBar and NA64.

Finally, we stress that the dark matter search strategy that was proposed here can be applied to many other experiments besides our benchmark choices, 
as long as they get the delayed neutrinos dominantly from non-relativistic muons induced by stopped-pion decays and timing of events is measured. 
We expect that such low-energy high-intensity neutrino facilities will make groundbreaking progress in the search for physics beyond the Standard Model.

\acknowledgments
We thank Pilar Coloma, Yuri Efremenko, Johnathon Jordan, Kevin Kelly, Soo-Bong Kim, Pedro Machado, Daniel Pershey, Rebecca Rapp, Grayson Rich, Rex Tayloe, Kate Scholberg, Richard Van de Water, and Jonghee Yoo for useful discussions. 
We specially appreciate Seongjin In for his \texttt{GEANT}4 simulations.
Portions of this research were conducted with the advanced computing resources provided by Texas A\&M High Performance Research Computing.
BD and LES acknowledge support from DOE Grant DE-SC0010813.
The work of DK is supported by the Department of Energy under Grant No. DE-FG02-13ER41976/DE-SC0009913.
SL acknowledges support from COS-STRP (TAMU). 
The work of JCP is supported by the National Research Foundation of Korea (NRF-2019R1C1C1005073 and NRF-2021R1A4A2001897). 
The work of SS is supported by the National Research Foundation of Korea
(NRF-2020R1I1A3072747).

\appendix
\section{Phase-Space Suppression of Cascade Photons \label{sec:appA}}

\begin{figure}
    \centering
    \includegraphics[width=0.65\linewidth]{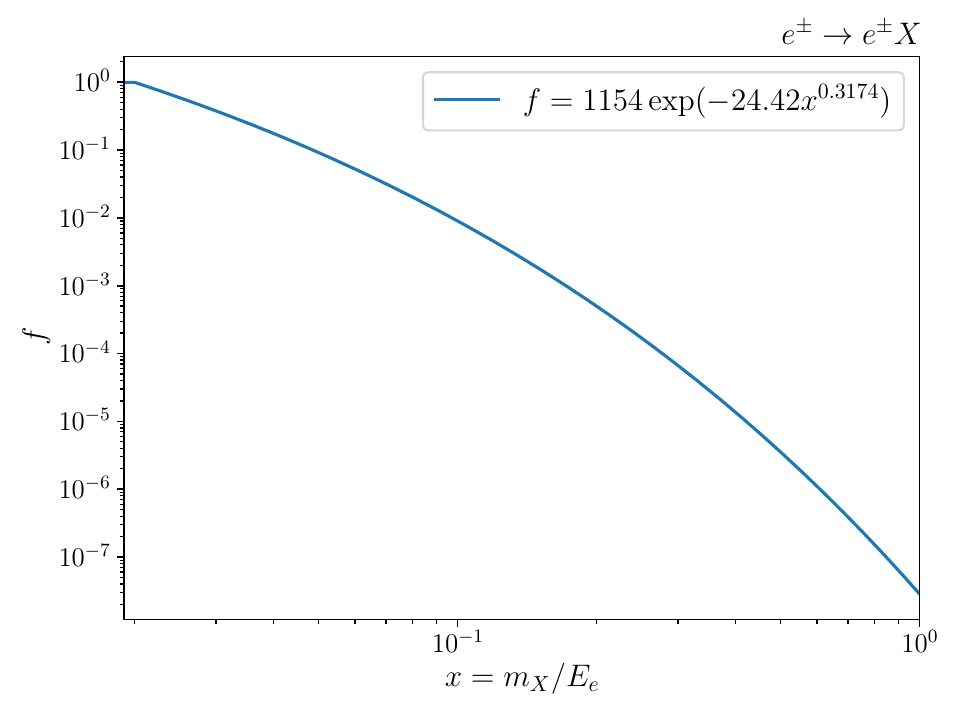}
    \caption{Effective phase-space factor suppression of dark photon production via $e^\pm$ cascade photons.}
    \label{fig:supp}
\end{figure}

We consider the dark bremsstrahlung process $$e^{\pm\,*} \rightarrow e^\pm \, X$$ for a dark gauge boson $X$ which has an interaction with electrons via $\mathcal{L} \supset \kappa_e^X x_e^X X_\mu \bar{e}\gamma^\mu e$ which is, in turn, contrasted with the ordinary QED, $\mathcal{L}_{\rm QED} \supset e A_\mu \bar{e}\gamma^\mu e$.
In a formal approach, one would supply the new physics model to an events generator, given $m_X$, $\kappa_e^X$, and the incident $e^\pm$ flux, and simulate this process.
However, this approach would be very computationally expensive, having to repeat it for each mass point in our likelihood analysis.
For the sake of pragmatism, at the cost of simulation fidelity, we instead parameterize the cross section as
\begin{equation}
\dfrac{d\sigma(e^\pm \to e^\pm X)}{dE_e} = \dfrac{d\sigma(e^\pm \to e^\pm \gamma)}{dE_e} \times p_X^2 \times f\bigg(\frac{m_X}{E_e}\bigg)
\end{equation}
where $f$ is a phase space factor that we obtain empirically from sampling several values of $m_X$ using \texttt{MG5@aMC}~\cite{Alwall:2014hca} and measuring the dependence on $m_X / E_e$.
We show the suppression $f$ as a function of $x = m_X / E_e$ in Figure~\ref{fig:supp}.
Furthermore, given a flux of $e^\pm$ cascade photons with energies $E_\gamma$, for each $\gamma$ we can approximately deduce the parent $e^\pm$ energy via
\begin{equation}
    \braket{E_e} = 1.0773 E_\gamma + 13.716 \, \,(\text{MeV}).
\end{equation}
This allows us to convolve an already-simulated, standard model photon flux with these factors to give the ad-hoc dark gauge boson flux for this process;
\begin{equation}
\dfrac{d\Phi_X}{dE_X} \sim \dfrac{d\Phi_\gamma}{dE_\gamma} \times p_X^2 \times f\bigg(\frac{m_X}{\braket{E_e}}\bigg).
\end{equation}

We find that our empirical model in Figure~\ref{fig:supp} works sufficiently well for $E_e>50$~MeV, whereas below 50~MeV the flux is additionally suppressed by $1-3$ orders of magnitude. 
However, light dark gauge boson $X$ contributed by the cascade photons from electrons of $E_e<50$~MeV can be produced more by $\pi^0$ decays and $\pi^-$ absorption.
Therefore, their contribution to our sensitivity reaches is subleading, i.e., the uncertainty in the associated $\Phi_X$ does not affect our sensitivity estimate. 

\section{Derivation of Timing Spectra \label{sec:appB}}

In this appendix, we derive the spectral shape of the timing distribution, taking the example of a simple, two-step, sequential $\pi^-$ absorption process defined by
\beq
\pi^- + p \to n+ X\,,~~~ X \to \chi\bar{\chi}\,.
\eeq
Such pions can be produced in all of our benchmark experiments, when a proton beam bombards on a target. 
The $\pi^-$ absorption and the emission of $X$ take place rather promptly inside the target, while the decay point of $X$ depends on its lifetime and velocity.

\begin{figure}[t]
\centering
\includegraphics[width=10cm]{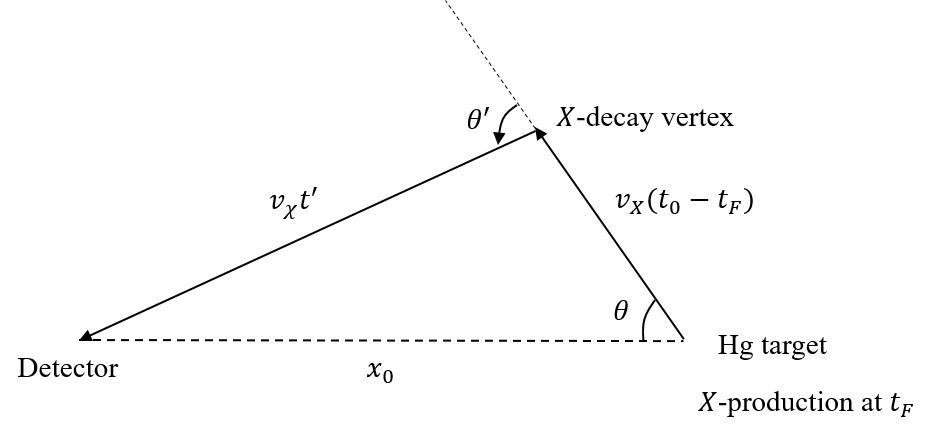}
\caption{\label{fig:config} The configuration under consideration.}
\end{figure}

The configuration under consideration is now depicted in Figure~\ref{fig:config}. 
A detector is placed in the origin, and the target is located at $x=x_0$. 
Suppose that the mesic state formed by a negative pion and a proton decays to dark gauge boson $X$ at $t_F$. 
As mentioned above, the formation of the mesic state followed by its decay proceed quickly, so one can understand $t_F$ as the timing of production of a given $\pi^-$. 
Suppose further that $X$ flies in the $\theta$ direction for $v_{X} (t_0-t_F)$, and decays to two dark matter particles. 
Since the mesic state is produced nearly at rest, the energy and the momentum of $X$ (denoted by $E_{X}$ and $p_{X}$, respectively) are given by the rest-frame values:
\begin{eqnarray}
E_{X} = \frac{(m_\pi+m_p)^2-m_n^2+m_{X}^2}{2(m_\pi+m_p)}\,,~~~p_{X} = \frac{\lambda^{1/2}\left[(m_\pi+m_p)^2,m_n^2,m_{X}^2\right]}{2(m_\pi+m_p)}\,,
\end{eqnarray}
where $m_i$ denotes the mass of particle species $i$. 
$v_{X}$, the speed of $X$ is trivially given by $p_{X}c/E_{X}$ with $c$ being the speed of light.
One of the two $\chi$ then may travel towards the detector for $v_\chi t'$, if it moves in the $\theta'$ direction (see also Figure~\ref{fig:config}).
The expression for $v_\chi$ is rather involved and, more importantly, dependent on time $(t_0-t_F)$, so we come back to it later.
Denoting the timing measured at the detector by $t$, we are interested in
\begin{equation}
f(t)=\frac{dN_\chi}{dt}\,,
\end{equation}
which can be interpreted as the dark matter flux at the detector of interest. 
Obviously, $t$ is the same as the sum of $t_0$ and $t'$:
\begin{equation}
t= t_0+t'\left(v_{X}(t_0-t_F), t_0-t_F,\cos\theta\right)\,,
\end{equation} 
where we emphasize that $t'$ is a function of $t_0-t_F$ and $\cos\theta$.

According to the decay law, the probability that $X$ decays at $t_0$ is given by $\frac{1}{\tau_{X}}e^{- (t_0-t_F)/\tau_{X}}$ with $\tau_{X}$ being the laboratory-frame mean lifetime of $X$. 
Assuming that the dark gauge boson is emitted isotropically in the process $\pi^- + p \to X + n$, we obtain
\begin{equation}
\frac{d^2N_{X}}{dt_0 d\cos\theta}=\frac{1}{2} \cdot \frac{1}{\tau_{X}}e^{-\frac{t_0-t_F}{\tau_{X}}}\Theta(t_0-t_F)\,,
\end{equation}
where $\Theta(x)$ is the usual Heaviside step function.
The differential dark matter number density is obviously proportional to the differential dark gauge boson number density, as it is from the decay of $X$. 
They are related by a simple change of variable such that
\begin{equation}
\frac{d^2 N_\chi}{dtd\cos\theta}\propto \left| \frac{\partial(t,\cos\theta)}{\partial(t_0,\cos\theta)} \right|^{-1}\frac{d^2N_{X}}{dt_0 d\cos\theta}=\left| \frac{dt}{dt_0} \right|^{-1}\frac{d^2N_{X}}{dt_0 d\cos\theta}\,.
\end{equation}
From Figure~\ref{fig:config} one can easily see that $t'$ is related to $t_0-t_F$ and $\cos\theta$ as follows: $(v_\chi t')^2 = x_0^2+v_{X}^2(t_0-t_F)^2-2x_0 v_{X}(t_0-t_F)\cos\theta$, which results in 
\begin{equation}
t = t_0+\frac{\sqrt{x_0^2+v_{X}^2(t_0-t_F)^2-2x_0 v_{X}(t_0-t_F)\cos\theta}}{v_\chi}\,.
\end{equation}
As mentioned earlier, $v_\chi$ is a function of $t_0$, thus the time-derivative of $t$ is rather involved and generally not illustrative. 
We provide example expressions for some limiting cases below.
Since $v_\chi$ depends on $t_0$, $dt/dt_0$ is
\begin{equation}
\frac{dt}{dt_0}=\frac{\partial t}{\partial t_0}+\frac{\partial v_\chi}{\partial t_0}\cdot \frac{\partial t}{\partial v_\chi}\,.
\end{equation}
In the limit of $m_\chi \ll m_{X}$, $v_\chi \to c$ and in turn $\partial v_\chi /\partial t_0=0$ so that we obtain
\begin{eqnarray}
\left|\frac{dt}{dt_0}\right|^{-1} &=& \frac{c \sqrt{x_0^2+v_{X}^2(t_0-t_F)^2-2x_0 v_{X}(t_0-t_F)\cos\theta}}{c \sqrt{x_0^2+v_{X}^2(t_0-t_F)^2-2x_0 v_{X}(t_0-t_F)\cos\theta}+v_{X}^2(t_0-t_F)-x_0 v_{X} \cos\theta}\,, \\
t_0 &=&t_F+\frac{1}{c^2-v_{X}^2}\left[c^2(t-t_F)-x_0v_{X}\cos\theta \right. \nonumber \\
&&\left. - \sqrt{c^2 v_{X}^2 (t-t_F)^2-2x_0v_{X} c^2(t-t_F)\cos\theta+x_0^2(c^2-v_{X}^2\sin^2\theta)}\right]\,.
\end{eqnarray}
In the limit of $m_\chi \ll m_{X} \ll m_\pi + m_p - m_n$, all the velocity parameters approach the speed of light so that the above expressions become further simplified.
\begin{eqnarray}
\left|\frac{dt}{dt_0}\right|^{-1} &=& \frac{\sqrt{x_0^2+c^2(t_0-t_F)^2-2x_0 c(t_0-t_F)\cos\theta}}{ \sqrt{x_0^2+c^2(t_0-t_F)^2-2x_0 c(t_0-t_F)\cos\theta}+c (t_0-t_F)-x_0 \cos\theta}\,, \\
t_0 &=&\frac{c^2(t^2-t_F^2)-2 c t_F x_0 \cos\theta -x_0^2 }{2c\{c(t-t_F)-x_0\cos\theta\}}\,.
\end{eqnarray}

As mentioned before, not all dark matter particles contribute to $f(t)$ but the ones traveling in the $\theta'$ direction do.
Therefore, such a contribution has to be properly weighted in terms of $\theta'$.  
To find the associated weight factor $w(\cos\theta')$, let us first suppose that starred quantities are measured in the $X$ rest frame. 
The Lorentz transformation of $\chi$ four-momentum between the laboratory frame and the $X$ rest frame leads us to the relation,
\begin{equation}
\cos\theta^* = \frac{-E_\chi^*\gamma_{X}\sqrt{\gamma_{X}^2-1}\sin^2\theta'-\sqrt{\cos^2\theta'\{E_\chi^{*2}-m_\chi^2(\cos^2\theta'+\gamma_{X}^2\sin^2\theta') \}}}{p_\chi^*(\cos^2\theta'+\gamma_{X}^2\sin^2\theta')}\,, \label{eq:coss}
\end{equation}
where $\gamma_{X}$ stands for the Lorentz boost factor of dark gauge boson.
Indeed, there is another solution in which the sign for the second term of the numerator is positive, but it does not describe the dark matter flying towards the detector.
Note that the emission direction of $\chi$ in the $X$ rest frame is isotropic, i.e.,
\begin{equation}
\frac{dN_{X\to\chi}}{d\cos\theta^*}=2\cdot \frac{1}{2}\,,
\end{equation}
where the prefactor 2 takes care of the fact that $X$ disintegrates to two dark matter particles. 
We therefore find that the weight factor $w$ is
\begin{equation}
w(\cos\theta')=\frac{1}{2\pi(v_\chi t')^2}\left|\frac{d\cos\theta'}{d\cos\theta^*} \right|^{-1}\frac{dN_{X\to\chi}}{d\cos\theta^*}
\end{equation}
where $(2\pi)^{-1}$ averages out the azimuthal angle around the axis defined by the $X$ moving orientation and $(v_\chi t')^{-2}$ takes care of the flux reduction by the distance between the $X$ decay point and the detector.
A simple geometry consideration relates $\cos\theta'$ and $\cos\theta$:
\begin{eqnarray}
\cos\theta'&=& \frac{x_0\cos\theta-v_{X}(t_0-t_F)}{\sqrt{x_0^2+v_{X}^2(t_0-t_F)^2-2x_0v_{X}(t_0-t_F)\cos\theta}}\,. \label{eq:cosp}
\end{eqnarray}
Now $v_\chi$ can be written in terms of $t$ and $\cos\theta$. 
\begin{equation}
v_\chi=\frac{p_\chi}{E_\chi}\cdot c=\frac{\sqrt{\left(E_\chi^*\gamma_X+p_\chi^*\cos\theta^* \sqrt{\gamma_X^2-1}\right)^2-m_\chi^2}}{E_\chi^*\gamma_X+p_\chi^*\cos\theta^* \sqrt{\gamma_X^2-1}}\cdot c
\end{equation}
One can re-express the above in terms of $\cos\theta$ and $t_0$ using Eqs.~\eqref{eq:coss} and \eqref{eq:cosp}.

Finally, the actual proton beam pulse is not maximized immediately, hence produced pion flux is not maximized instantaneously. It actually rises for a certain amount of time, culminates, and falls off. Suppose that such a beam pulse is on for $t_F^{\max}$ and its behavior in the timing spectrum is modeled by unit-normalized $\mathcal{F}$. 
Since production of negative pion is proportional to protons on target, we finally find
\begin{equation}
\frac{dN_\chi}{dt}=\int_{-1}^{1} d\cos\theta \int_{0}^{t_F^{\max}} dt_F \left| \frac{dt}{dt_0}\right|^{-1} \frac{d^2 N_X}{dt_0 d\cos\theta} \cdot w(\cos\theta')\cdot \mathcal{F}(t_F)\,,
\end{equation}
where every quantity is written in terms of $t$.

\bibliographystyle{JHEP}
\bibliography{main}

\end{document}